\documentclass[prl, twocolumn, superscriptaddress]{revtex4-1}
\usepackage{bm, amsmath, amsfonts, amssymb, braket}
\usepackage{times}
\usepackage{multirow}
\usepackage[dvipdfmx]{graphicx}
\usepackage{float}
\usepackage{color}
\usepackage{comment}

\begin{document}

\newcommand{\ii}{\text{i}}
\newcommand{\Z}{$\mathbb{Z}$}
\newcommand{\Zt}{$\mathbb{Z}_{2}$}

\newcommand{\PT}{{\cal PT}}
\newcommand{\CP}{{\cal CP}}
\newcommand{\Tp}{{\cal T}_{+}}
\newcommand{\Tm}{{\cal T}_{-}}
\newcommand{\Cp}{{\cal C}_{+}}
\newcommand{\Cm}{{\cal C}_{-}}
\newcommand{\CS}{\Gamma}
\newcommand{\SLS}{{\cal S}}
\newcommand{\Lr}{~$\text{L}_{\rm r}$}
\newcommand{\Li}{~$\text{L}_{\rm i}$}

\newcommand{\Ca}{${\cal C}_{p}$ & \Z & $0$ & \Z & $0$ & \Z & $0$ & \Z & $0$}
\newcommand{\Caa}{${\cal C}_{p} \times {\cal C}_{p}$ & $\mathbb{Z} \oplus \mathbb{Z}$ & $0$ & $\mathbb{Z} \oplus \mathbb{Z}$ & $0$ & $\mathbb{Z} \oplus \mathbb{Z}$ & $0$ & $\mathbb{Z} \oplus \mathbb{Z}$ & $0$}

\newcommand{\Cb}{${\cal C}_{p+1}$ & $0$ & \Z & $0$ & \Z & $0$ & \Z & $0$ & \Z}
\newcommand{\Cbb}{${\cal C}_{p+1} \times {\cal C}_{p+1}$ & $0$ & $\mathbb{Z} \oplus \mathbb{Z}$ & $0$ & $\mathbb{Z} \oplus \mathbb{Z}$ & $0$ & $\mathbb{Z} \oplus \mathbb{Z}$ & $0$ & $\mathbb{Z} \oplus \mathbb{Z}$}

\newcommand{\Ra}{${\cal R}_{p}$ & \Z & \Zt & \Zt & $0$ & 2\Z & $0$ & $0$ & $0$}
\newcommand{\Raa}{${\cal R}_{p} \times {\cal R}_{p}$ & $\mathbb{Z} \oplus \mathbb{Z}$ & $\mathbb{Z}_{2} \oplus \mathbb{Z}_{2}$ & $\mathbb{Z}_{2} \oplus \mathbb{Z}_{2}$ & $0$ & $2\mathbb{Z} \oplus 2\mathbb{Z}$ & $0$ & $0$ & $0$}

\newcommand{\Rb}{${\cal R}_{p+1}$ & \Zt & \Zt & $0$ & 2\Z & $0$ & $0$ & $0$ & \Z}
\newcommand{\Rbb}{${\cal R}_{p+1} \times {\cal R}_{p+1}$ & $\mathbb{Z}_{2} \oplus \mathbb{Z}_{2}$ & $\mathbb{Z}_{2} \oplus \mathbb{Z}_{2}$ & $0$ & $2\mathbb{Z} \oplus 2\mathbb{Z}$ & $0$ & $0$ & $0$ & $\mathbb{Z} \oplus \mathbb{Z}$}

\newcommand{\Rc}{${\cal R}_{p+2}$ & \Zt & $0$ & 2\Z & $0$ & $0$ & $0$ & \Z & \Zt}
\newcommand{\Rcc}{${\cal R}_{p+2} \times {\cal R}_{p+2}$ & $\mathbb{Z}_{2} \oplus \mathbb{Z}_{2}$ & $0$ & $2\mathbb{Z} \oplus 2\mathbb{Z}$ & $0$ & $0$ & $0$ & $\mathbb{Z} \oplus \mathbb{Z}$ & $\mathbb{Z}_{2} \oplus \mathbb{Z}_{2}$}

\newcommand{\Rd}{${\cal R}_{p+3}$ & $0$ & 2\Z & $0$ & $0$ & $0$ & \Z & \Zt & \Zt}
\newcommand{\Rdd}{${\cal R}_{p+3} \times {\cal R}_{p+3}$ & $0$ & $2\mathbb{Z} \oplus 2\mathbb{Z}$ & $0$ & $0$ & $0$ & $\mathbb{Z} \oplus \mathbb{Z}$ & $\mathbb{Z}_{2} \oplus \mathbb{Z}_{2}$ & $\mathbb{Z}_{2} \oplus \mathbb{Z}_{2}$}

\newcommand{\Ree}{${\cal R}_{p+4}$ & 2\Z & $0$ & $0$ & $0$ & \Z & \Zt & \Zt & $0$}
\newcommand{\Reee}{${\cal R}_{p+4} \times {\cal R}_{p+4}$ & $2\mathbb{Z} \oplus 2\mathbb{Z}$ & $0$ & $0$ & $0$ & $\mathbb{Z} \oplus \mathbb{Z}$ & $\mathbb{Z}_{2} \oplus \mathbb{Z}_{2}$ & $\mathbb{Z}_{2} \oplus \mathbb{Z}_{2}$ & $0$}

\newcommand{\Rf}{${\cal R}_{p+5}$ & $0$ & $0$ & $0$ & \Z & \Zt & \Zt & $0$ & 2\Z}
\newcommand{\Rff}{${\cal R}_{p+5} \times {\cal R}_{p+5}$ & $0$ & $0$ & $0$ & $\mathbb{Z} \oplus \mathbb{Z}$ & $\mathbb{Z}_{2} \oplus \mathbb{Z}_{2}$ & $\mathbb{Z}_{2} \oplus \mathbb{Z}_{2}$ & $0$ & $2\mathbb{Z} \oplus 2\mathbb{Z}$}

\newcommand{\Rg}{${\cal R}_{p+6}$ & $0$ & $0$ & \Z & \Zt & \Zt & $0$ & 2\Z & $0$}
\newcommand{\Rgg}{${\cal R}_{p+6} \times {\cal R}_{p+6}$ & $0$ & $0$ & $\mathbb{Z} \oplus \mathbb{Z}$ & $\mathbb{Z}_{2} \oplus \mathbb{Z}_{2}$ & $\mathbb{Z}_{2} \oplus \mathbb{Z}_{2}$ & $0$ & $2\mathbb{Z} \oplus 2\mathbb{Z}$ & $0$}

\newcommand{\Rh}{${\cal R}_{p+7}$ & $0$ & \Z & \Zt & \Zt & $0$ & 2\Z & $0$ & $0$}
\newcommand{\Rhh}{${\cal R}_{p+7} \times {\cal R}_{p+7}$ & $0$ & $\mathbb{Z} \oplus \mathbb{Z}$ & $\mathbb{Z}_{2} \oplus \mathbb{Z}_{2}$ & $\mathbb{Z}_{2} \oplus \mathbb{Z}_{2}$ & $0$ & $2\mathbb{Z} \oplus 2\mathbb{Z}$ & $0$ & $0$}

\title{Classification of Exceptional Points and Non-Hermitian Topological Semimetals}

\author{Kohei Kawabata}
\email{kawabata@cat.phys.s.u-tokyo.ac.jp}
\affiliation{Department of Physics, University of Tokyo, 7-3-1 Hongo, Bunkyo-ku, Tokyo 113-0033, Japan}

\author{Takumi Bessho}
\email{takumi.bessho@yukawa.kyoto-u.ac.jp}
\affiliation{Yukawa Institute for Theoretical Physics, Kyoto University, Kyoto 606-8502, Japan}

\author{Masatoshi Sato}
\email{msato@yukawa.kyoto-u.ac.jp}
\affiliation{Yukawa Institute for Theoretical Physics, Kyoto University, Kyoto 606-8502, Japan}

\date{\today}

\begin{abstract}
Exceptional points are universal level degeneracies induced by non-Hermiticity. Whereas past decades witnessed their new physics, the unified understanding has yet to be obtained. Here we present the complete classification of generic topologically stable exceptional points according to two types of complex-energy gaps and fundamental symmetries of charge conjugation, parity, and time reversal. This classification reveals unique non-Hermitian gapless structures with no Hermitian analogs and systematically predicts unknown non-Hermitian semimetals and nodal superconductors; a {\it topological dumbbell} of exceptional points in three dimensions is constructed as an illustration. Our work paves the way toward richer phenomena and functionalities of exceptional points and non-Hermitian topological semimetals.
\end{abstract}

\maketitle

%%%%% Introduction %%%%%
Topology plays a pivotal role in the understanding of phases of matter~\cite{Kane-review, *Zhang-review, *Schnyder-Ryu-review}. In gapless systems such as semimetals and nodal superconductors, topology guarantees stable degeneracies accompanying distinctive excitations~\cite{NN-83, Murakami-07, Wan-11, BB-11, *BHB-11, *HB-12, *MB-12, Young-12, Armitage-review}. A prime example is the Weyl semimetal in three dimensions, where each gapless point is topologically protected by the Chern number defined on the enclosing surface. Symmetry further brings about diverse types of topological semimetals. Their unified understanding is developed as the classification theory according to fundamental symmetries such as $\mathcal{PT}$ and $\mathcal{CP}$ symmetries~\cite{Horava-05, Matsuura-13, Zhao-13, *Zhao-14, Morimoto-14, Kobayashi-14, Shiozaki-14, Chiu-14, Zhao-16, *Zhao-17}. 

Recently, the interplay between topology and non-Hermiticity has attracted widespread interest in a non-Hermitian extension of topological insulators~\cite{Rudner-09, *Zeuner-15, Hu-11, Esaki-11, *Sato-12, Poli-15, Malzard-15, Lee-16, Weimann-17, Leykam-17, Obuse-17, *Xiao-17, St-Jean-17, Bahari-17, Harari-18, *Bandres-18, Shen-18, Kunst-18, KAKU-18, YW-18, *YSW-18, Lieu-18, Gong-18, Philip-18, *Hirsbrunner-19, KSU-18, Takata-18, Clerk-18, KHGAU-19, Bliokh-19, Liu-19, Edvardsson-19, Lee-19, Herviou-19, Kunst-19, KSUS-18, ZL-19, Zeng-19, Zirnstein-19, Rui-19, Borgnia-19, Yokomizo-19, McClarty-19, Longhi-19} and semimetals~\cite{Zhen-15, Gonzalez-16, *Gonzalez-17, *Molina-18, Xu-17, Chernodub-17, *Chernodub-19, Zyuzin-18, *Moors-19, *Zyuzin-19, Cerjan-18, Zhou-18-exp, Carlstrom-18, *Carlstrom-19, Okugawa-19, Budich-19, Yang-19, Zhou-19, Wang-19, Yoshida-19, Kozii-17, *Papaj-19, *Shen-Fu-18, Cerjan-18-exp, Luo-18, Lee-18-tidal, Bergholtz-19}. Non-Hermiticity ubiquitously appears, for instance, in nonequilibrium open systems~\cite{Konotop-review, *Feng-review, *Christodoulides-review, *Alu-review} and correlated electron systems~\cite{Kozii-17, *Papaj-19, *Shen-Fu-18}, leading to unusual properties with no Hermitian counterparts. One of their salient characteristics is the emergence of exceptional points~\cite{Kato, Berry-04, Heiss-12}, i.e., universal non-Hermitian level degeneracies at which eigenstates coalesce with and linearly depend on each other~\cite{defective}. The past decade has witnessed a plethora of rich phenomena and functionalities induced by exceptional points~\cite{Konotop-review, *Feng-review, *Christodoulides-review, *Alu-review}, including unidirectional invisibility~\cite{Lin-11, Regensburger-12, Feng-13, Peng-14}, chiral transport~\cite{Dembowski-01, Gao-15, Doppler-16, Xu-16, Yoon-18}, enhanced sensitivity~\cite{Wiersig-14, Liu-16, Hodaei-17, Chen-17, Lau-18, Zhang-18}, and unusual quantum criticality~\cite{Lee-14, KAU-17, Nakagawa-18, Li-19, Dora-19, Wu-19, Xiao-18}.

Such exceptional points also alter the nodal structures of topological semimetals in a fundamental manner~\cite{Zhen-15, Gonzalez-16, *Gonzalez-17, *Molina-18, Xu-17, Chernodub-17, *Chernodub-19, Zyuzin-18, *Moors-19, *Zyuzin-19, Cerjan-18, Zhou-18-exp, Carlstrom-18, *Carlstrom-19, Okugawa-19, Budich-19, Yang-19, Zhou-19, Wang-19, Yoshida-19, Kozii-17, *Papaj-19, *Shen-Fu-18, Cerjan-18-exp, Luo-18, Lee-18-tidal, Bergholtz-19}. Notably, non-Hermiticity deforms a Weyl point and spawns a ring of exceptional points~\cite{Berry-04}. This Weyl exceptional ring is characterized by two topological charges~\cite{Xu-17}, a Chern number and a quantized Berry phase, and such a multiple topological structure has no analogs to Weyl and Dirac points in Hermitian systems~\cite{NN-83, Murakami-07, Wan-11, BB-11, *BHB-11, *HB-12, *MB-12, Young-12, Horava-05, Matsuura-13, Zhao-13, *Zhao-14, Morimoto-14, Kobayashi-14, Shiozaki-14, Chiu-14, Zhao-16, *Zhao-17, Armitage-review}. Its experimental observation has also been reported in an optical waveguide array~\cite{Cerjan-18-exp}. Moreover, pseudo-Hermiticity~\cite{Budich-19}, $\mathcal{PT}$ symmetry~\cite{Okugawa-19, Zhou-19}, and chiral symmetry~\cite{Yoshida-19} enable an exceptional ring (surface) in two (three) dimensions. A symmetry-protected exceptional ring has been experimentally observed in a two-dimensional photonic crystal slab with $\mathcal{PT}$ symmetry~\cite{Zhen-15}. Whereas these unconventional nodal structures imply a fundamental change in the existing classification for Hermitian topological semimetals, its non-Hermitian counterpart has yet to be established. The non-Hermitian topological classification is crucial not only because it provides a general theoretical framework but also because it predicts novel non-Hermitian topological materials and serves as a benchmark for experiments.

\begin{figure}[t]
\centering
\includegraphics[width=86mm]{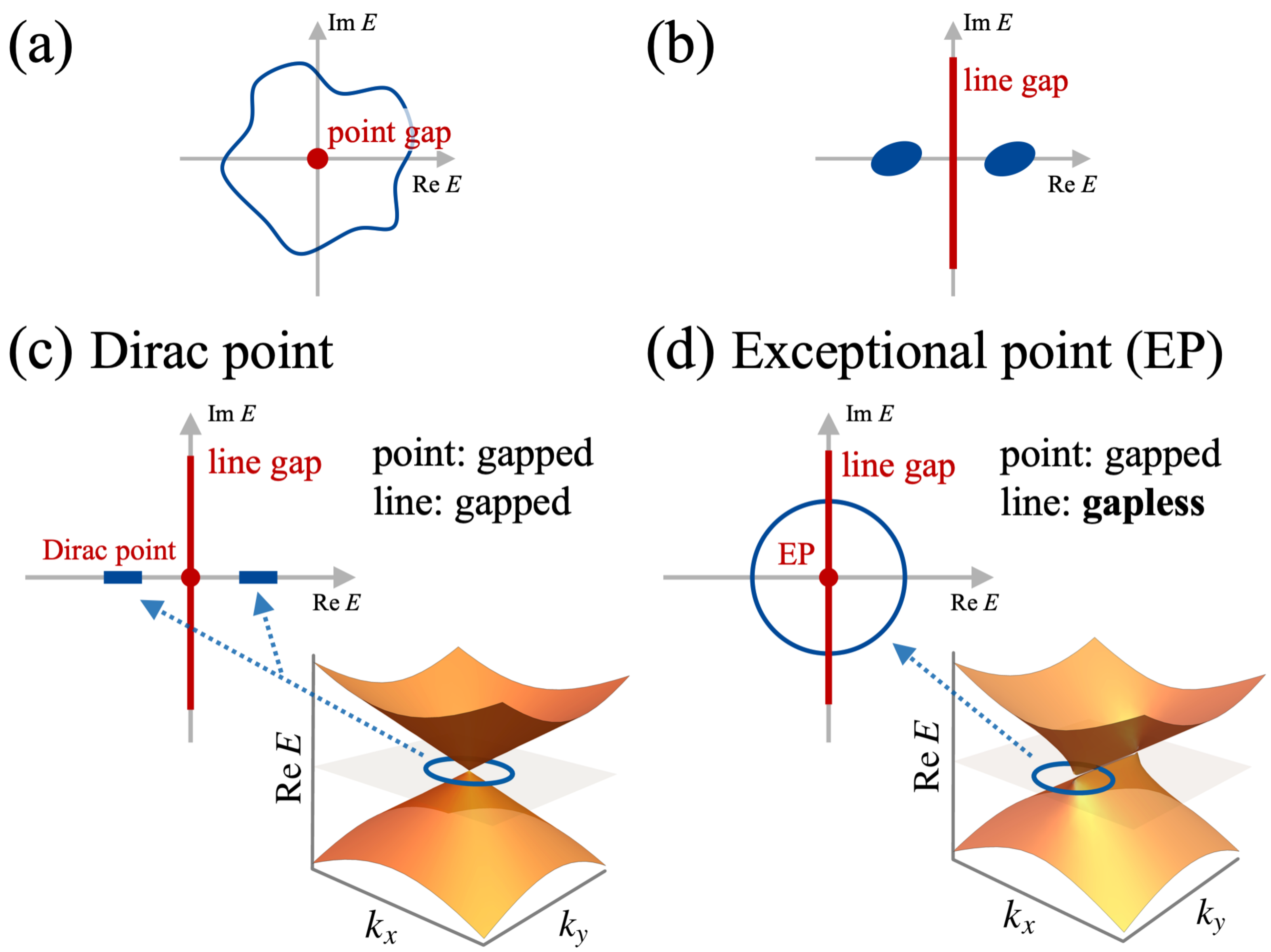} 
\caption{Non-Hermitian gapless structures. Complex spectra (blue regions) in non-Hermitian systems may host two types of energy gaps: (a)~point gap and (b)~line gap. A point (line) gap is open when the complex spectrum does not cross a reference point (line) in the complex-energy plane. (c)~Hermitian gapless point (Dirac point). On a region (blue circle) around it, both point and line gaps are open. (d)~Exceptional point. On a region (blue circle) around it, a point gap is open but a line gap is closed.}
	\label{fig: gapless}
\end{figure}

\begin{figure}[t]
\centering
\includegraphics[width=86mm]{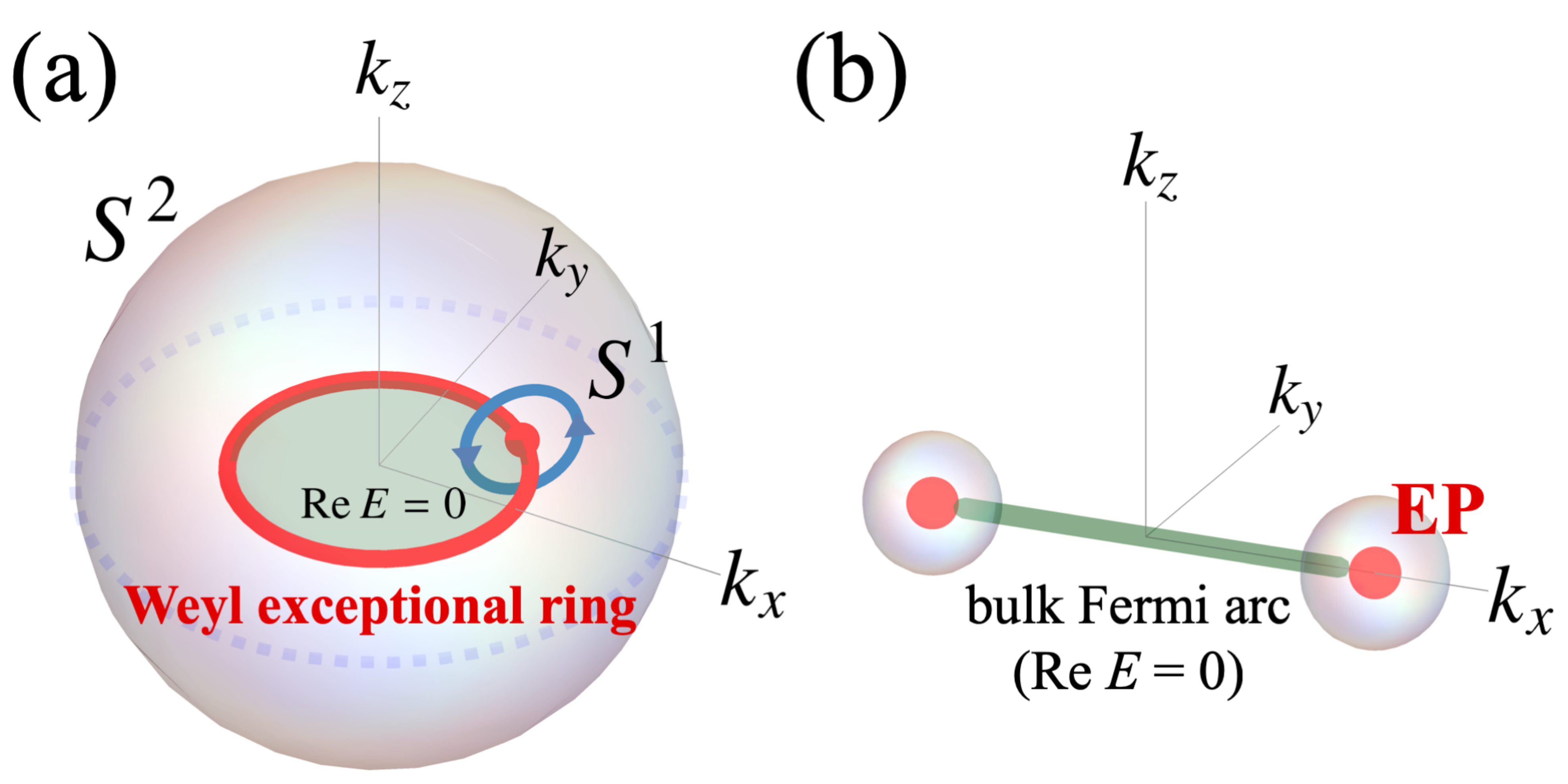} 
\caption{Non-Hermitian topological semimetals. (a)~Weyl exceptional ring (red ring) with the two independent topological charges, the Chern number defined on the surface $S^{2}$ that encloses the ring ($p=3$, line gap) and the winding number defined on the loop $S^{1}$ across the ring ($p=2$, point gap). (b)~Topological dumbbell of exceptional points in three dimensions. The bulk Fermi arc with $\mathrm{Re}\,E=0$ connects a pair of exceptional points (EPs, red points).}
	\label{fig: NH-TSM}
\end{figure}

This Letter presents the general theory that completely classifies topologically stable exceptional points at generic momentum points in non-Hermitian topological semimetals according to fundamental 38-fold symmetry and two types of complex-energy gaps (Table~\ref{tab: classification}~\cite{supplement}). Our classification is based on the observation that non-Hermiticity enables a unique gapless structure with no Hermitian analogs: only one type of the complex-energy gaps is open around an exceptional point [Fig.~\ref{fig: gapless}\,(d)], which is sharply contrasted with conventional Weyl and Dirac points in Hermitian systems around which both types are open [Fig.~\ref{fig: gapless}\,(c)]. We also elucidate that exceptional points generally possess multiple topological structures due to the two types of complex-energy gaps. Our theory provides the unified understanding of non-Hermitian topological semimetals studied in previous works~\cite{Zhen-15, Gonzalez-16, *Gonzalez-17, *Molina-18, Xu-17, Chernodub-17, *Chernodub-19, Zyuzin-18, *Moors-19, *Zyuzin-19, Cerjan-18, Zhou-18-exp, Carlstrom-18, *Carlstrom-19, Okugawa-19, Budich-19, Yang-19, Zhou-19, Wang-19, Yoshida-19, Kozii-17, *Papaj-19, *Shen-Fu-18, Cerjan-18-exp, Luo-18, Lee-18-tidal, Bergholtz-19}. Furthermore, it systematically predicts novel non-Hermitian topological materials unnoticed in the literature; we construct as an illustration a topological dumbbell of exceptional points [Fig.~\ref{fig: NH-TSM}\,(b)], i.e., a three-dimensional variant of the bulk Fermi arcs~\cite{Zhou-18-exp, Kozii-17, *Papaj-19, *Shen-Fu-18}.

%%%%%%%%%%
\paragraph{Non-Hermitian gapless structures.\,---}

A unique characteristic of non-Hermitian systems is complex-valued eigenenergy, from which two different energy gaps are defined, a point gap [Fig.~\ref{fig: gapless}\,(a)] and a line gap [Fig.~\ref{fig: gapless}\,(b)]~\cite{KSUS-18}. In the presence of a point (line) gap, complex-energy bands do not cross a reference point (line) in the complex-energy plane. A point gap is physically relevant to the localization transition in one-dimensional non-Hermitian systems~\cite{Hatano-96, *Hatano-97, *Hatano-98, Gong-18, Longhi-19}, whereas topologically protected edge states~\cite{Esaki-11, *Sato-12, Poli-15, Weimann-17, Obuse-17, *Xiao-17, St-Jean-17, Bahari-17, Harari-18, *Bandres-18, Shen-18} are understood by a line gap for the real part of the complex spectrum. A line gap for the imaginary part also has a significant influence on the nonequilibrium dynamics~\cite{KHGAU-19}. If symmetry exists, both complex-energy gaps should be invariant under the symmetry. Without loss of generality, the reference point is supposed to be placed on the reference line. Then, a line gap is always closed when a point gap is closed. However, the converse is not necessarily true; a point gap can be open even when a line gap is closed.

This nature of complex-energy gaps enables two distinct types of non-Hermitian gapless structures. We encircle a gapless region (point, line, surface, and so on) in momentum space by a $(p-1)$-dimensional sphere $S^{p-1}$ ($p \geq 1$), where $S^{0}$, $S^{1}$, and $S^{2}$ denote a pair of points, a circle, and a surface, respectively. The system has a complex-energy gap on $S^{p-1}$, but two different situations may happen: (i)~both point and line gaps are open [Fig.~\ref{fig: gapless}\,(c)] or (ii)~only a point gap is open [Fig.~\ref{fig: gapless}\,(d)]. In the former case, the non-Hermitian Hamiltonian on $S^{p-1}$ can be continuously deformed into a Hermitian (or an anti-Hermitian) one as rigorously proven in Ref.~\cite{KSUS-18}. Thus, in a manner analogous to the Chern number for the conventional Weyl point, the gapless region hosts a topological charge essentially identical to that in the Hermitian case. In the latter case, by contrast, we need to assign a different topological charge to the gapless region on the basis of the point gap on $S^{p-1}$. It should be noted that the latter is intrinsic to non-Hermitian systems and impossible in Hermitian ones, since there is no distinction between point and line gaps for Hermitian Hamiltonians.

We find that exceptional points realize the latter unique gapless structure as shown in Fig.~\ref{fig: gapless}\,(d) and are characterized by topological charges for point gaps. A distinctive property of an exceptional point is swapping of eigenenergies and eigenstates upon its encirclement~\cite{Heiss-12, Konotop-review, *Feng-review, *Christodoulides-review, *Alu-review, Dembowski-01, Gao-15, Doppler-16, Xu-16, Zhou-18-exp, Yoon-18}. For illustration, let us consider a two-dimensional system with no symmetry that has an exceptional point ${\bm k} = {\bm k}_{\rm EP}$ at which two complex bands $E_{\pm} \left( {\bm k} \right)$ coalesce. A representative model is given as $H \left( {\bm k} \right) = k_{x} \sigma_{x} + \left( k_{y} + \ii \gamma \right) \sigma_{y}$ with Pauli matrices $\sigma_{x, y}$ and the degree of non-Hermiticity $\gamma$. The eigenenergies are $E_{\pm} \left( {\bm k} \right) = \pm \sqrt{{\bm k}^{2} - \gamma^{2} + 2\ii \gamma k_{y}}$ with the square root singularity around the exceptional points ${\bm k}_{\rm EP} = \left( \pm \gamma, 0 \right)$. As a direct result of this singularity, a branch cut and a self-intersecting Riemann surface appear in the complex-energy plane and thus $E_{+} \left( {\bm k} \right)$ and $E_{-} \left( {\bm k} \right)$ are swapped when we go around a loop $S^{1}$ in momentum space that encircles one of the exceptional points. Importantly, a point gap for the reference point $E \left( {\bm k}_{\rm EP} \right)$ is open but a line gap is closed on $S^{1}$. As a result, the determinant of $H \left( \bm{k} \right) - E \left( {\bm k}_{\rm EP} \right)$ does not vanish on $S^{1}$, which enables us to define the winding number for a point gap as
\begin{equation}
W := \oint_{S^{1}} \frac{d{\bm k}}{2\pi \ii} \cdot \nabla_{\bm k} \log \det \left[ H \left( {\bm k} \right) - E \left( {\bm k}_{\rm EP} \right) \right].
	\label{eq: A p=2}
\end{equation}
An exceptional point with $W \neq 0$ is topologically stable. For the above representative model, the exceptional points ${\bm k}_{\rm EP} = \left( \pm \gamma, 0 \right)$ are characterized by $W = \pm 1$, respectively. We note that a similar topological invariant $\nu$ called vorticity is introduced in Refs.~\cite{Leykam-17, Shen-18}: $\nu := - \left( 2\pi \right)^{-1} \oint_{S^{1}} d{\bm k} \cdot \nabla_{\bm k}\,\mathrm{arg} \left[ E_{+} \left( {\bm k} \right) - E_{-} \left( {\bm k} \right) \right]$. Whereas both invariants are equivalent to each other for the above simple two-band model ($\nu = -W/2$), only $W$ is straightforwardly applicable to general models at which more than two complex bands coalesce. 

It is shown that exceptional points that accompany spontaneous $\PT$-symmetry breaking~\cite{Heiss-12, Konotop-review, *Feng-review, *Christodoulides-review, *Alu-review, Bender-98, *Bender-02, *Bender-review} also possess a similar gapless structure (see Sec.~SV in Ref.~\cite{supplement}): only a point gap is open and a line gap is closed around them.

%%%%% classification table %%%%%
\begin{table*}[t]
	\centering
	\caption{Classification table of topologically stable exceptional points at generic momentum points. The codimension $p$ is defined as $p := d-d_{\rm EP}$ with the spatial dimension $d$ and the dimension $d_{\rm EP}$ of the gapless region; exceptional points, lines, and surfaces are described by $d_{\rm EP} = 0, 1, 2$, respectively. Complex-energy gaps have two distinct types, a point (P) or line (L) gap, and the subscript of L specifies a line gap for the real or imaginary part of the complex spectrum. The sign of $\PT$ ($\CP$) symmetry means $(\PT)\,(\PT)^{*}$ [$(\CP)\,(\CP)^{*}$].}
	\label{tab: classification}
     \begin{tabular}{ccccccccccc} \hline \hline
    ~~Symmetry~~ & ~Gap~ & ~Classifying space~ & ~~$p=0$~~ & ~~$p=1$~~ & ~~$p=2$~~ & ~~$p=3$~~ & ~~$p=4$~~ & ~~$p=5$~~ & ~~$p=6$~~ & ~~$p=7$~~ \\ \hline
    \multirow{2}{*}{No} 
    & P & \Ca \\ 
    & L & \Cb \\ \hline
    \multirow{3}{*}{Chiral} 
    & P & \Cb \\ 
    & \Lr & \Ca \\
    & \Li & \Cbb \\ \hline
    \multirow{2}{*}{Sublattice} 
    & P & \Caa \\ 
    & L & \Ca \\ \hline
    \multirow{3}{*}{$\PT$, $+1$} 
    & P & \Ra \\ 
    & \Lr & \Rh \\
    & \Li & \Rb \\ \hline
    \multirow{3}{*}{$\PT$, $-1$} 
    & P & \Ree \\ 
    & \Lr & \Rd \\
    & \Li & \Rf \\ \hline
    \multirow{2}{*}{$\CP$, $+1$} 
    & P & \Rc \\ 
    & L & \Rb \\ \hline
    \multirow{2}{*}{$\CP$, $-1$} 
    & P & \Rg \\ 
    & L & \Rf \\ \hline \hline
  \end{tabular}
\end{table*}
%%%%%%%%%%

%%%%%%%%%%
\paragraph{Symmetry.\,---}
To classify topologically stable exceptional points at generic momentum points for both complex-energy gaps in a general manner, we consider fundamental symmetries that keep all momenta invariant. The prime examples are $\PT$ and $\CP$ symmetries defined by
\begin{eqnarray}
(\PT)\,H^{*} \left( {\bm k} \right) (\PT)^{-1} &=& H \left( {\bm k} \right),~
(\PT) (\PT)^{*} = \pm 1,\quad	
	\label{eq: PT symmetry} \\
(\CP)\,H^{T} \left( {\bm k} \right) (\CP)^{-1} &=& - H \left( {\bm k} \right),~
(\CP) (\CP)^{*} = \pm 1,\quad 
	 \label{eq: CP symmetry}
\end{eqnarray}
where $\PT$ and $\CP$ are unitary operators. Here $\PT$ symmetry emerges in open systems with balanced gain and loss~\cite{Konotop-review, *Feng-review, *Christodoulides-review, *Alu-review}, while $\CP$ symmetry is respected in non-Hermitian superconductors and superfluids with inversion symmetry~\cite{PHS}. Whereas both space inversion ${\cal P}$ and time reversal ${\cal T}$ (charge conjugation ${\cal C}$) may be broken individually, the combined symmetry $\PT$ ($\CP$) is required for the topological stability away from high-symmetry points. 

Another relevant symmetry is chiral symmetry, which is a combination of $\PT$ and $\CP$ symmetries and hence is defined by
\begin{eqnarray}
\Gamma H^{\dag} \left( {\bm k} \right) \Gamma^{-1} = - H \left( {\bm k} \right),\quad
\Gamma^{2} = 1,
	\label{eq: chiral symmetry}
\end{eqnarray}
with a unitary operator $\Gamma$. It is notable that chiral symmetry is equivalent to pseudo-Hermiticity, defined by $\eta H^{\dag} \left( \bm k \right) \eta^{-1} = H \left( \bm k \right)$ with a unitary and Hermitian operator $\eta$~\cite{Mostafazadeh-02-1, *Mostafazadeh-02-2, *Mostafazadeh-02-3}. In fact, when $H \left( {\bm k} \right)$ respects chiral symmetry, $\ii H \left( {\bm k} \right)$ respects pseudo-Hermiticity~\cite{KHGAU-19}. Moreover, chiral symmetry is distinct from sublattice symmetry due to $H \left( {\bm k} \right) \neq H^{\dag} \left( {\bm k} \right)$~\cite{KSUS-18}, defined by 
\begin{eqnarray}
{\cal S} H \left( {\bm k} \right) {\cal S}^{-1} = - H \left( {\bm k} \right),\quad
{\cal S}^{2} = 1,
	\label{eq: sublattice symmetry}
\end{eqnarray}
with a unitary operator ${\cal S}$. The above symmetries constitute the fundamental 38-fold symmetry class for the non-Hermitian gapless phases~\cite{supplement} in a similar manner to the gapped ones in Ref.~\cite{KSUS-18}, which updates Bernard-LeClair symmetry~\cite{Bernard-LeClair-02, Esaki-11, *Sato-12, Lieu-18, Budich-19, ZL-19, 38}.

%%%%%%%%%%
\paragraph{Topological classification.\,---}

Complex-energy gaps are closed at an exceptional point. Near such a gapless point, $H \left( {\bm k} \right)$ is expressed as~\cite{Bloch} 
\begin{equation}
H \left( {\bm k} \right)
= \sum_{i=1}^{p} v_{i} \delta k_{i} \gamma_{i} + PJP^{-1},\quad
p := d-d_{\rm EP},
	\label{eq: Hamiltonian; near-gapless}
\end{equation}
where $v_{i}$'s are complex constants, $\delta k_{i}$'s are momentum deviations from the exceptional point, and $\gamma_{i}$'s are non-Hermitian Dirac matrices subject to certain symmetries. The gapless point $\delta {\bm k} = 0$ in general takes a Jordan canonical form $PJP^{-1}$ with a Jordan matrix $J$ and an invertible matrix $P$, which is defective unless $J$ is diagonal~\cite{Kato}. In addition, $p$ is the codimension of the exceptional point with the spatial dimension $d$ and the dimension $d_{\rm EP}$ of the gapless region; exceptional points, lines, and surfaces are described by $d_{\rm EP} = 0, 1, 2$, respectively. Whereas a complex-energy gap is closed at the gapless point $\delta {\bm k} = 0$, it may be open, and a topological invariant can be defined on a ($p-1$)-dimensional surface $S^{p-1}$ that encloses the gapless point, in a similar manner to Fermi surfaces in Hermitian systems~\cite{Morimoto-14, Kobayashi-14, Zhao-16, Karoubi, Schnyder-Ryu-review}. Thus, topology of the exceptional point is determined by classifying the gapped topological phases on $S^{p-1}$. 

As discussed above, only a point gap is open for a sufficiently small $S^{p-1}$, but a line gap can also be open for a larger $S^{p'-1}$ [$p'$ can be different from $p$; for instance, $p=2$ and $p'=3$ in Fig.~\ref{fig: NH-TSM}\,(a)]. Depending on types of the complex-energy gaps, an exceptional point may support multiple topological invariants. Taking into account both possibilities, we classify topologically stable exceptional points for each type of complex-energy gaps and all of the 38 symmetry classes~\cite{supplement}. Our results are summarized in Table~\ref{tab: classification} and Tables~S2-S7 in Supplemental Material~\cite{supplement}. These periodic tables specify exceptional points and non-Hermitian topological semimetals in a general manner and describe their unconventional nodal structures. In fact, they corroborate previous works~\cite{Zhen-15, Gonzalez-16, *Gonzalez-17, *Molina-18, Xu-17, Chernodub-17, *Chernodub-19, Zyuzin-18, *Moors-19, *Zyuzin-19, Cerjan-18, Zhou-18-exp, Carlstrom-18, *Carlstrom-19, Okugawa-19, Budich-19, Yang-19, Zhou-19, Wang-19, Yoshida-19, Kozii-17, *Papaj-19, *Shen-Fu-18, Cerjan-18-exp, Luo-18, Lee-18-tidal, Bergholtz-19, supplement}. For example, stable exceptional points in two dimensions~\cite{Shen-18, Zhou-18-exp, Kozii-17, *Papaj-19, *Shen-Fu-18, Bergholtz-19} are explained by the $\mathbb{Z}$ index in Table~\ref{tab: classification} with no symmetry, $p=2$, and point (P) gap; symmetry-protected exceptional rings in two dimensions~\cite{Zhen-15, Okugawa-19, Budich-19, Yoshida-19, Zhou-19} are explained by the $\mathbb{Z}$ or $\mathbb{Z}_{2}$ index in Table~\ref{tab: classification} with chiral or $\PT$ symmetry, $p=1$, and point (P) gap. These unique nodal structures, as well as the consequent physical phenomena, are topologically protected against symmetry-preserving perturbations.

%%%%%%%%%%
\paragraph{Multiple topological structure.\,---}

A remarkable feature of Table~\ref{tab: classification} is the multiple topological indices for each symmetry class due to the two types of complex-energy gaps. As a result, exceptional points can be characterized by a couple of independent topological charges. A prime example is the Weyl exceptional ring in three-dimensional systems with no symmetry [Fig.~\ref{fig: NH-TSM}\,(a)]~\cite{Berry-04, Xu-17}. For the Hermitian case, the topological stability of a Weyl point in three dimensions is ensured by the Chern number defined on an enclosing surface~\cite{NN-83, Armitage-review}. In the presence of non-Hermiticity, such a Weyl point morphs into an exceptional ring. A representative model is given as $H \left( {\bm k} \right) = k_{x} \sigma_{x} + k_{y} \sigma_{y} + \left( k_{z} + \ii \gamma \right) \sigma_{z}$ with a Weyl exceptional ring at $k_{x}^{2} + k_{y}^{2} = \gamma^{2}$, $k_{z} = 0$~\cite{Berry-04, Xu-17}. On a sphere [$S^{2}$ in Fig.~\ref{fig: NH-TSM}\,(a)] that encloses the ring, a line gap is open and the Chern number remains to be well defined unless it annihilates with another ring, which corresponds to the $\mathbb{Z}$ index for a line gap with $p=3$ (see Table~\ref{tab: classification} with no symmetry). Moreover, on $S^{1}$ across the ring, a point gap is open and the winding number in Eq.~(\ref{eq: A p=2}) is well defined, which corresponds to the $\mathbb{Z}$ index for a point gap with $p=2$. Notably, the two topological charges are independent of each other and individually ensure the topological stability of the Weyl exceptional ring.

Such a multiple topological structure is a general hallmark of non-Hermitian topological semimetals. For instance, according to our classification tables, symmetry-protected exceptional rings in two dimensions~\cite{Zhen-15, Okugawa-19, Budich-19, Zhou-19, Yoshida-19} may host different topological charges for each $p \leq 2$ and each complex-energy gap. Whereas the unique nodal structures were discussed in Refs.~\cite{Zhen-15, Okugawa-19, Budich-19, Yoshida-19, Zhou-19}, their multiple topology has not been revealed until this Letter.

%%%%% chiral, p=3 %%%%%
\paragraph{Exceptional point in three dimensions.\,---}
Our classification also predicts unknown non-Hermitian topological semimetals and nodal superconductors. As an illustrative example, we consider an exceptional point in three dimensions ($p=3$) protected by chiral symmetry. Whereas exceptional rings~\cite{Berry-04, Xu-17} and surfaces~\cite{Okugawa-19, Zhou-19} were discussed before, such an exceptional point in three dimensions has not hitherto been known. A representative model is systematically constructed in the following manner. We begin with a Hermitian gapless system in four dimensions $H \left( {\bm k} \right) = k_{x} \sigma_{x} \tau_{x} + k_{y} \sigma_{x} \tau_{y} + k_{z} \sigma_{x} \tau_{z} + k_{w} \sigma_{y}$ with Pauli matrices $\sigma_{i}$'s and $\tau_{i}$'s. Corresponding to the $\mathbb{Z}$ index for a real line gap with $p=4$ (see Table~\ref{tab: classification} with chiral symmetry), it possesses a topologically stable gapless point at ${\bm k} = 0$, around which the three-dimensional winding number~\cite{supplement, Schnyder-08, *Ryu-10, Qi-10, Kawakami-18} is defined due to chiral symmetry in Eq.~(\ref{eq: chiral symmetry}) with $\Gamma = \sigma_{z}$. Now we add a non-Hermitian perturbation $\ii \gamma \sigma_{z} \tau_{x}$ to this Hermitian model. As with the Weyl exceptional ring, this non-Hermiticity spawns an exceptional ring. In fact, the complex spectrum is obtained as
\begin{equation} 
E \left( {\bm k} \right) = \pm \sqrt{k_{x}^{2} + k_{w}^{2} + ( \sqrt{k_{y}^{2} + k_{z}^{2}} \pm \ii \gamma )^{2} },
\end{equation} 
and an exceptional ring appears at $k_{x}^{2} + k_{w}^{2} = \gamma^{2}$, $k_{y} = k_{z} = 0$. Finally, we take $k_{w} = 0$ and regard this four-dimensional model as a three-dimensional one,
\begin{equation}
H \left( {\bm k} \right) = k_{x} \sigma_{x} \tau_{x} + k_{y} \sigma_{x} \tau_{y} + k_{z} \sigma_{x} \tau_{z} + \ii \gamma \sigma_{z} \tau_{x},
	\label{eq: CS-p3}
\end{equation}
which has a pair of exceptional points at ${\bm k}_{\rm EP} = \left( \pm \gamma, 0, 0 \right)$. As illustrated in Fig.~\ref{fig: NH-TSM}\,(b), these exceptional points are connected by a Fermi arc with $\mathrm{Re}\,E = 0$, forming a topologically stable dumbbell configuration. The topological stability of the exceptional points is ensured by the $\mathbb{Z}$ index in Table~\ref{tab: classification} with chiral symmetry, $p=3$, and point gap, which is given as the Chern number $\pm 1$ of the Hermitian matrix $\ii H \left( {\bm k} \right) \Gamma$ defined on a surface that encloses each of the exceptional points~\cite{supplement}.

We stress that topologically stable gapless points are absent in Hermitian three-dimensional systems with chiral symmetry. These gapless points are thus unique to non-Hermitian systems. Furthermore, the above recipe is widely applicable to different symmetry classes and spatial dimensions and unknown non-Hermitian topological semimetals can be systematically predicted according to our classification.

%%%%% Conclusion %%%%%
\paragraph{Discussion.\,---}
The emergence of exceptional points is one of the most striking and universal characteristics in non-Hermitian physics~\cite{Berry-04, Heiss-12, Konotop-review, *Feng-review, *Christodoulides-review, *Alu-review}. In this Letter, we have formulated the general classification theory on non-Hermitian topology of exceptional points. Our theory is widely applicable to non-Hermitian topological semimetals, providing the unified understanding of their unique nodal structures and predicting novel non-Hermitian topological materials. Since the consequent non-Hermitian topological phenomena and functionalities have yet to be explored, it merits further research to investigate such richer physics on the basis of our theory.

%%%%% Acknowledgement %%%%%
We are grateful to Ken Shiozaki and Masahito Ueda for the valuable discussions. K.K. also thanks Emil J. Bergholtz, Ryo Okugawa, and Tsuneya Yoshida for the helpful discussions. This work was supported by a Grant-in-Aid for Scientific Research on Innovative Areas ``Topological Materials Science" (KAKENHI Grant No.~JP15H05855) from the Japan Society for the Promotion of Science (JSPS). K.K. was supported by the JSPS through the Program for Leading Graduate Schools (ALPS) and KAKENHI Grant No.~JP19J21927, and in part through the domestic junior researcher exchange program of ``Topological Materials Science." M.S. was supported by KAKENHI Grant No.~JP17H02922 from the JSPS.

K.K. and T.B. contributed equally to this work.

\bibliography{NH-EP-TSM}

\begin{thebibliography}{142}
\bibitem [{def()}]{defective}%
  \BibitemOpen
  \href@noop {} {}\bibinfo {note} {{In Ref.~\cite{Kato}, an exceptional point
  is defined as a generic degeneracy point at which the matrix may be
  diagonalizable. On the other hand, Ref.~\cite{Kato} defines a degeneracy
  point of a nondiagonalizable matrix as a defective point.}}\BibitemShut
  {Stop}%
\bibitem [{PHS()}]{PHS}%
  \BibitemOpen
  \href@noop {} {}\bibinfo {note} {{Particle-hole symmetry in non-Hermitian
  superconductors and superfluids should be defined in terms not of complex
  conjugation but of transposition~\cite{KSUS-18}.}}\BibitemShut {Stop}%
\bibitem [{38()}]{38}%
  \BibitemOpen
  \href@noop {} {}\bibinfo {note} {{Reference~\cite{KSUS-18} points out that
  the 43-fold Bernard-LeClair symmetry overcounted some non-Hermitian symmetry
  classes and overlooked others. Consequently, the non-Hermitian generalization
  of the Altland-Zirnbauer symmetry actually consists of 38
  classes~\cite{KSUS-18}. Furthermore, the physical meaning of each symmetry is
  identified in Ref.~\cite{KSUS-18}.}}\BibitemShut {Stop}%
\bibitem [{Blo()}]{Bloch}%
  \BibitemOpen
  \href@noop {} {}\bibinfo {note} {Our discussion is applicable to the
  generalized Bloch wave functions~\cite{YW-18, *YSW-18, Kunst-19, KSUS-18,
  Yokomizo-19} as long as the corresponding symmetry is respected with complex
  wavenumbers.}\BibitemShut {Stop}%
\end{thebibliography}

%%%%% Supplemental Material %%%%%
\widetext
\pagebreak

\renewcommand{\theequation}{S\arabic{equation}}
\renewcommand{\thefigure}{S\arabic{figure}}
\renewcommand{\thetable}{S\arabic{table}}
\setcounter{equation}{0}
\setcounter{figure}{0}
\setcounter{table}{0}

\begin{center}
{\bf \large Supplemental Material for \\ \smallskip ``Classification of Exceptional Points and Non-Hermitian Topological Semimetals"}
\end{center}

%%%%%%%%%%
\section{SI.~Symmetry}

\begin{table}[b]
	\centering
	\caption{${\cal P}$AZ and ${\cal P} \text{AZ}^{\dag}$ symmetry classes for non-Hermitian Hamiltonians. $\mathcal{P}\mathcal{T}_{+}$ symmetry, $\mathcal{C}_{-}\mathcal{P}$ symmetry, and chiral symmetry are respectively defined by $({\cal P}{\cal T}_{+})\,H^{*} \left( {\bm k} \right) ({\cal P}{\cal T}_{+})^{-1} = H \left( {\bm k} \right)$ with $({\cal P}{\cal T}_{+}) ({\cal P}{\cal T}_{+})^{*} = \pm 1$, $({\cal C}_{-}{\cal P})\,H^{T} \left( {\bm k} \right) ({\cal C}_{-}{\cal P})^{-1} = - H \left( {\bm k} \right)$ with $({\cal C}_{-}{\cal P}) ({\cal C}_{-}{\cal P})^{*} = \pm 1$, and $\Gamma\,H^{\dag} \left( {\bm k} \right) \Gamma^{-1} = - H \left( {\bm k} \right)$ with $\Gamma^{2} = 1$, which constitute ${\cal P}$AZ symmetry class. $\mathcal{C}_{+}\mathcal{P}$ symmetry and $\mathcal{P}\mathcal{T}_{-}$ symmetry are respectively defined by $({\cal C}_{+}{\cal P})\,H^{T} \left( {\bm k} \right) ({\cal C}_{+}{\cal P})^{-1} = H \left( {\bm k} \right)$ with $({\cal C}_{+}{\cal P}) ({\cal C}_{+}{\cal P})^{*} = \pm 1$ and $({\cal P}{\cal T}_{-})\,H^{*} \left( {\bm k} \right) ({\cal P}{\cal T}_{-})^{-1} = - H \left( {\bm k} \right)$ with $({\cal P}{\cal T}_{-}) ({\cal P}{\cal T}_{-})^{*} = \pm 1$, which constitute ${\cal P} \text{AZ}^{\dag}$ symmetry class.\\}
		\label{tab: AZ}
     \begin{tabular}{ccccccc} \hline \hline
    \multicolumn{2}{c}{~~~Symmetry class~~~} & ~~~$\mathcal{P}\mathcal{T}_{+}$~~~ & ~~~${\cal C}_{-}{\cal P}$~~~ & ~~~${\cal C}_{+}{\cal P}$~~~ & ~~~$\mathcal{P}\mathcal{T}_{-}$~~~ & ~~~$\Gamma$~~~ \\ \hline
    \multirow{2}{*}{~Complex AZ~} 
    & A & $0$ & $0$ & $0$ & $0$ & $0$ \\
    & AIII & $0$ & $0$ & $0$ & $0$ & $1$ \\ \hline
    \multirow{9}{*}{Real ${\cal P}$AZ} 
    & ${\cal P}$AI & $+1$ & $0$ & $0$ & $0$ & $0$ \\
    & ${\cal P}$BDI & $+1$ & $+1$ & $0$ & $0$ & $1$ \\
    & ${\cal P}$D & $0$ & $+1$ & $0$ & $0$ & $0$ \\
    & ~${\cal P}$DIII~ & $-1$ & $+1$ & $0$ & $0$ & $1$ \\
    & ${\cal P}$AII & $-1$ & $0$ & $0$ & $0$ & $0$ \\
    & ${\cal P}$CII & $-1$ & $-1$ & $0$ & $0$ & $1$ \\
    & ${\cal P}$C & $0$ & $-1$ & $0$ & $0$ & $0$ \\
    & ${\cal P}$CI & $+1$ & $-1$ & $0$ & $0$ & $1$ \\ \hline
    \multirow{9}{*}{Real ${\cal P}\text{AZ}^{\dag}$} 
    & ${\cal P}\text{AI}^{\dag}$ & $0$ & $0$ & $+1$ & $0$ & $0$ \\
    & ${\cal P}\text{BDI}^{\dag}$ & $0$ & $0$ & $+1$ & $+1$ & $1$ \\
    & ${\cal P}\text{D}^{\dag}$ & $0$ & $0$ & $0$ & $+1$ & $0$ \\
    & ${\cal P}\text{DIII}^{\dag}$ & $0$ & $0$ & $-1$ & $+1$ & $1$ \\
    & ${\cal P}\text{AII}^{\dag}$ & $0$ & $0$ & $-1$ & $0$ & $0$ \\
    & ${\cal P}\text{CII}^{\dag}$ & $0$ & $0$ & $-1$ & $-1$ & $1$ \\
    & ${\cal P}\text{C}^{\dag}$ & $0$ & $0$ & $0$ & $-1$ & $0$ \\
    & ${\cal P}\text{CI}^{\dag}$ & $0$ & $0$ & $+1$ & $-1$ & $1$ \\ \hline \hline
    \end{tabular}
\end{table} 

To investigate topologically stable exceptional points that appear at generic momentum points, we consider the following 38-fold symmetry~\cite{KSUS-18} that keeps all momenta invariant. First of all, $\PT$ symmetry in Eq.~(\ref{eq: PT symmetry}), $\CP$ symmetry in Eq.~(\ref{eq: CP symmetry}), and chiral symmetry in Eq.~(\ref{eq: chiral symmetry}) constitute the 10-fold symmetry class, which is denoted as parity-equipped Altland-Zirnbauer ($\mathcal{P}$AZ) symmetry class (Table~\ref{tab: AZ}). In stark contrast to Hermitian systems, a Hermitian-conjugate counterpart of $\mathcal{P}$AZ symmetry can be introduced independently. To distinguish between them, we denote $\PT$ ($\CP$) symmetry in the main text as $\mathcal{P}\mathcal{T}_{+}$ ($\mathcal{C}_{-}\mathcal{P}$) symmetry in Table~\ref{tab: AZ}. The Hermitian-conjugate counterpart of $\mathcal{P}\mathcal{T}_{+}$ symmetry is defined with transposition instead of complex conjugation by
\begin{equation}
(\mathcal{C}_{+}\mathcal{P})\,H^{T} \left( {\bm k} \right) (\mathcal{C}_{+}\mathcal{P})^{-1}
= H \left( {\bm k} \right),\quad
(\mathcal{C}_{+}\mathcal{P}) (\mathcal{C}_{+}\mathcal{P})^{*} = \pm 1,
\end{equation}
with a unitary operator $\mathcal{C}_{+}\mathcal{P}$, and the Hermitian-conjugate counterpart of $\mathcal{C}_{-}\mathcal{P}$ symmetry is defined with complex conjugation instead of transposition by
\begin{equation}
(\mathcal{P}\mathcal{T}_{-})\,H^{*} \left( {\bm k} \right) (\mathcal{P}\mathcal{T}_{-})^{-1}
= - H \left( {\bm k} \right),\quad
(\mathcal{P}\mathcal{T}_{-}) (\mathcal{P}\mathcal{T}_{-})^{*} = \pm 1,
\end{equation}
with a unitary operator $\mathcal{P}\mathcal{T}_{-}$. In our notation, $\mathcal{C}_{\pm}$ and $\mathcal{T}_{\pm}$ imply transposition and complex conjugation, respectively. Importantly, $\mathcal{P}\mathcal{T}_{+}$ and $\mathcal{C}_{-}\mathcal{P}$ symmetries respectively coincide with $\mathcal{C}_{+}\mathcal{P}$ and $\mathcal{P}\mathcal{T}_{-}$ symmetries for Hermitian Hamiltonians [$H \left( {\bm k} \right) = H^{\dag} \left( {\bm k} \right)$], but they do not in the presence of non-Hermiticity since complex conjugation and transposition are not equivalent by definition [$H^{*} \left( {\bm k} \right) \neq H^{T} \left( {\bm k} \right)$]. Here $\mathcal{C}_{+}\mathcal{P}$ and $\mathcal{P}\mathcal{T}_{-}$ symmetries in addition to chiral symmetry also constitute the 10-fold symmetry class, which is denoted as $\mathcal{P}\text{AZ}^{\dag}$ symmetry class (Table~\ref{tab: AZ}). Here we abbreviate the Hermitian-conjugate counterpart of $\mathcal{P}$AZ symmetry class as $\mathcal{P}\text{AZ}^{\dag}$ symmetry class (similar abbreviations are used below). It should be noted that all the symmetry classes in Table~\ref{tab: AZ} are not independent. Complex AZ class and its Hermitian conjugate (i.e., complex $\text{AZ}^{\dag}$ class) are equivalent to each other. Class ${\cal P}$AI (${\cal P}$AII) is equivalent to class ${\cal P}\text{D}^{\dag}$ (${\cal P}\text{C}^{\dag}$). In fact, when a non-Hermitian Hamiltonian $H \left( {\bm k} \right)$ belongs to class ${\cal P}$AI (${\cal P}$AII), another non-Hermitian Hamiltonian $\ii H \left( {\bm k} \right)$ belongs to class ${\cal P}\text{D}^{\dag}$ (${\cal P}\text{C}^{\dag}$)~\cite{KHGAU-19}. In this manner, we totally have 16 independent symmetry classes in Table~\ref{tab: AZ}.

Since sublattice symmetry in Eq.~(\ref{eq: sublattice symmetry}) is distinct from chiral symmetry in Eq.~(\ref{eq: chiral symmetry}) for non-Hermitian Hamiltonians, it can be considered as an additional symmetry to ${\cal P}$AZ symmetry. Specifying the commutation or anticommutation relation between $\mathcal{S}$ and the other symmetry operators, we have 3 (19) additional symmetry classes for the complex AZ (real ${\cal P}$AZ) class with sublattice symmetry~\cite{KSUS-18}. We can also add sublattice symmetry to ${\cal P} \text{AZ}^{\dag}$ symmetry class, but each real ${\cal P} \text{AZ}^{\dag}$ class with sublattice symmetry is equivalent to the corresponding real ${\cal P}$AZ class with sublattice symmetry. Whereas pseudo-Hermiticity can also be considered as an additional symmetry to ${\cal P}$AZ symmetry, each ${\cal P}$AZ or ${\cal P} \text{AZ}^{\dag}$ symmetry class with pseudo-Hermiticity is equivalent to the corresponding ${\cal P}$AZ or ${\cal P} \text{AZ}^{\dag}$ symmetry class, or that with sublattice symmetry. In total, we have $16 + 3 + 19 = 38$ symmetry classes.

%%%%%%%%%%
\section{SII.~Classification tables}

%%%%% AZ %%%%%
\begin{table}[b]
	\centering
	\caption{Topological classification table of exceptional points in ${\cal P}$AZ symmetry class. Exceptional points are classified according to ${\cal P}$AZ symmetry class, the codimension $p$, and the definition of a complex-energy point (P) or line (L) gap. The codimension $p$ is defined as $p := d-d_{\rm EP}$ with the spatial dimension $d$ and the dimension $d_{\rm EP}$ of the exceptional point. The subscript of L specifies the line gap for the real or imaginary part of the complex spectrum.\\}
	\label{tab: complex + real AZ}
     \begin{tabular}{ccccccccccc} \hline \hline
    ~${\cal P}\text{AZ}$ class~ & ~Gap~ & Classifying space & ~$p=0$~ & ~$p=1$~ & ~$p=2$~ & ~$p=3$~& ~$p=4$~ & ~$p=5$~ & ~$p=6$~ & ~$p=7$~ \\ \hline
    \multirow{2}{*}{A} 
    & P & \Ca \\ 
    & L & \Cb \\ \hline
    \multirow{3}{*}{AIII} 
    & P & \Cb \\ 
    & \Lr & \Ca \\
    & \Li & \Cbb \\ \hline \hline
    \multirow{3}{*}{${\cal P}$AI} 
    & P & \Ra \\ 
    & \Lr & \Rh\\
    & \Li & \Rb \\ \hline
    \multirow{3}{*}{${\cal P}$BDI} 
    & P & \Rb \\ 
    & \Lr & \Ra \\
    & \Li & \Rbb \\ \hline
    \multirow{2}{*}{${\cal P}$D} 
    & P & \Rc \\ 
    & L & \Rb \\ \hline
    \multirow{3}{*}{${\cal P}$DIII} 
    & P & \Rd \\ 
    & \Lr & \Rc \\
    & \Li & \Cb \\ \hline
    \multirow{3}{*}{${\cal P}$AII} 
    & P & \Ree \\ 
    & \Lr & \Rd \\
    & \Li & \Rf \\ \hline
    \multirow{3}{*}{${\cal P}$CII} 
    & P & \Rf \\ 
    & \Lr & \Ree \\
    & \Li & \Rff \\ \hline
    \multirow{2}{*}{${\cal P}$C} 
    & P & \Rg \\ 
    & L & \Rf \\ \hline
    \multirow{3}{*}{${\cal P}$CI} 
    & P & \Rh \\ 
    & \Lr & \Rg \\
    & \Li & \Cb \\ \hline \hline
  \end{tabular}
\end{table}

In Tables~\ref{tab: complex + real AZ}-\ref{tab: real AZ + pH}, we provide topological classification of exceptional points at generic momentum points for all the 38 symmetry classes and two types of complex-energy gaps. A part of these tables is shown in Table~\ref{tab: classification} in the main text. Detailed derivations of these classification tables are provided in Sec.~SIII. We note that each symmetry class in Tables~\ref{tab: complex AZ + pH} and \ref{tab: real AZ + pH} is equivalent to the corresponding symmetry class in Tables~\ref{tab: complex + real AZ}-\ref{tab: real AZ + SLS}~\cite{KSUS-18}. Moreover, the following seven pairs of the symmetry classes in Tables~\ref{tab: complex + real AZ}-\ref{tab: real AZ + SLS} are equivalent to each other, respectively: (${\cal P}$AI \& ${\cal P}\text{D}^{\dag}$), (${\cal P}$AII \& ${\cal P}\text{C}^{\dag}$), (${\cal P}$AI + ${\cal S}_{-}$ \& ${\cal P}$AII + ${\cal S}_{-}$), (${\cal P}$BDI + ${\cal S}_{-+}$ \& ${\cal P}$DIII + ${\cal S}_{-+}$), (${\cal P}$BDI + ${\cal S}_{--}$ \& ${\cal P}$DIII + ${\cal S}_{--}$), (${\cal P}$CII + ${\cal S}_{-+}$ \& ${\cal P}$CI + ${\cal S}_{-+}$), and (${\cal P}$CII + ${\cal S}_{--}$ \& ${\cal P}$CII + ${\cal S}_{-+}$). Here ${\cal P}$AI + ${\cal S}_{-}$ denotes class ${\cal P}$AI with additional sublattice symmetry ${\cal S}_{-}$ that anticommutes with $\mathcal{P}\mathcal{T}_{+}$ symmetry, and so on.

%%%%% real AZ^{\dag} %%%%%
\begin{table}[H]
	\centering
	\caption{Topological classification table of exceptional points in real ${\cal P}\text{AZ}^{\dag}$ symmetry class. Exceptional points are classified according to ${\cal P}\text{AZ}^{\dag}$ symmetry class, the codimension $p$, and the definition of a complex-energy point (P) or line (L) gap. The codimension $p$ is defined as $p := d-d_{\rm EP}$ with the spatial dimension $d$ and the dimension $d_{\rm EP}$ of the exceptional point. The subscript of L specifies the line gap for the real or imaginary part of the complex spectrum.\\}
     \begin{tabular}{ccccccccccc} \hline \hline
    ~${\cal P}\text{AZ}^{\dag}$ class~ & ~Gap~ & Classifying space & ~$p=0$~ & ~$p=1$~ & ~$p=2$~ & ~$p=3$~& ~$p=4$~ & ~$p=5$~ & ~$p=6$~ & ~$p=7$~ \\ \hline
    \multirow{2}{*}{${\cal P}\text{AI}^{\dag}$} 
    & P & \Rg \\ 
    & L & \Rh \\ \hline
    \multirow{3}{*}{${\cal P}\text{BDI}^{\dag}$} 
    & P & \Rh \\ 
    & \Lr & \Ra \\
    & \Li & \Rhh \\ \hline
    \multirow{3}{*}{${\cal P}\text{D}^{\dag}$} 
    & P & \Ra \\
    & \Lr & \Rb \\ 
    & \Li & \Rh \\ \hline
    \multirow{3}{*}{${\cal P}\text{DIII}^{\dag}$} 
    & P & \Rb \\ 
    & \Lr & \Rc \\
    & \Li & \Cb \\ \hline
    \multirow{2}{*}{${\cal P}\text{AII}^{\dag}$} 
    & P & \Rc \\ 
    & L & \Rd \\ \hline
    \multirow{3}{*}{${\cal P}\text{CII}^{\dag}$} 
    & P & \Rd \\ 
    & \Lr & \Ree \\
    & \Li & \Rdd \\ \hline
    \multirow{3}{*}{${\cal P}\text{C}^{\dag}$} 
    & P & \Ree \\ 
    & \Lr & \Rf \\
    & \Li & \Rd \\ \hline
    \multirow{3}{*}{${\cal P}\text{CI}^{\dag}$} 
    & P & \Rf \\ 
    & \Lr & \Rg \\
    & \Li & \Cb \\ \hline \hline
  \end{tabular}
\end{table}
%%%%%%%%%%

%%%%% Complex AZ + SLS %%%%%
\begin{table}[H]
	\centering
	\caption{Topological classification table of exceptional points in complex AZ symmetry class with sublattice symmetry (SLS) ${\cal S}$. Exceptional points are classified according to the symmetry class, the codimension $p$, and the definition of a complex-energy point (P) or line (L) gap. The subscript of ${\cal S}$ specifies the commutation ($+$) or anticommutation ($-$) relation to chiral symmetry. The codimension $p$ is defined as $p := d-d_{\rm EP}$ with the spatial dimension $d$ and the dimension $d_{\rm EP}$ of the exceptional point. The subscript of L specifies the line gap for the real or imaginary part of the complex spectrum.\\}
     \begin{tabular}{cccccccccccc} \hline \hline
    ~SLS~ & ~AZ class~ & ~Gap~ & Classifying space & ~$p=0$~ & ~$p=1$~ & ~$p=2$~ & ~$p=3$~& ~$p=4$~ & ~$p=5$~ & ~$p=6$~ & ~$p=7$~ \\ \hline
	\multirow{3}{*}{$\SLS_{+}$} & \multirow{3}{*}{AIII}
	& P & \Ca \\
	& & \Lr & \Caa \\
	& & \Li & \Caa \\ \hline \hline
	\multirow{2}{*}{$\SLS$} & \multirow{2}{*}{A}
	& P & \Caa \\
	& & L & \Ca \\ \hline
	\multirow{3}{*}{$\SLS_{-}$} & \multirow{3}{*}{AIII}
	& P & \Cbb \\
	& & \Lr & \Cb \\
	& & \Li & \Cb \\ \hline \hline
  \end{tabular}
\end{table}
%%%%%%%%%%

%%%%% Real AZ + SLS %%%%%
\begin{table}[H]
	\centering
	\caption{Topological classification table of exceptional points in real ${\cal P}$AZ symmetry class with sublattice symmetry (SLS) ${\cal S}$. Exceptional points are classified according to the symmetry class, the codimension $p$, and the definition of a complex-energy point (P) or line (L) gap. The subscript of ${\cal S}$ specifies the commutation ($+$) or anticommutation ($-$) relation to $\mathcal{P}\mathcal{T}_{+}$ symmetry and/or $\mathcal{C}_{-}\mathcal{P}$ symmetry; for the symmetry classes that involve both $\mathcal{P}\mathcal{T}_{+}$ and $\mathcal{C}_{-}\mathcal{P}$ symmetries (${\cal P}$BDI, ${\cal P}$DIII, ${\cal P}$CII, and ${\cal P}$CI), the first and second subscripts specify the relation to $\mathcal{P}\mathcal{T}_{+}$ and $\mathcal{C}_{-}\mathcal{P}$ symmetries, respectively. The codimension $p$ is defined as $p := d-d_{\rm EP}$ with the spatial dimension $d$ and the dimension $d_{\rm EP}$ of the exceptional point. The subscript of L specifies the line gap for the real or imaginary part of the complex spectrum.\\}
	\label{tab: real AZ + SLS}
     \begin{tabular}{cccccccccccc} \hline \hline
    ~SLS~ & ~${\cal P}$AZ class~ & ~Gap~ & Classifying space & ~$p=0$~ & ~$p=1$~ & ~$p=2$~ & ~$p=3$~& ~$p=4$~ & ~$p=5$~ & ~$p=6$~ & ~$p=7$~ \\ \hline
	\multirow{3}{*}{$\SLS_{++}$} & \multirow{3}{*}{${\cal P}$BDI}
	& P & \Ra \\
	& & \Lr & \Raa \\
	& & \Li & \Raa \\ \hline
	\multirow{3}{*}{$\SLS_{--}$} & \multirow{3}{*}{${\cal P}$DIII}
	& P & \Rc \\
	& & \Lr & \Rcc \\
	& & \Li & \Ca \\ \hline
	\multirow{3}{*}{$\SLS_{++}$} & \multirow{3}{*}{${\cal P}$CII}
	& P & \Ree \\
	& & \Lr & \Reee \\
	& & \Li & \Reee \\ \hline
	\multirow{3}{*}{$\SLS_{--}$} & \multirow{3}{*}{${\cal P}$CI}
	& P & \Rg \\
	& & \Lr & \Rgg \\
	& & \Li & \Ca \\ \hline \hline
	\multirow{3}{*}{$\SLS_{-}$} & \multirow{3}{*}{${\cal P}$AI}
	& P & \Ca \\
	& & \Lr & \Rg \\
	& & \Li & \Rc \\ \hline
	\multirow{3}{*}{$\SLS_{-+}$} & \multirow{3}{*}{${\cal P}$BDI}
	& P & \Cb \\
	& & \Lr & \Rh \\
	& & \Li & \Rb \\ \hline
	\multirow{2}{*}{$\SLS_{+}$} & \multirow{2}{*}{${\cal P}$D}
	& P & \Ca \\
	& & L & \Ra \\ \hline
	\multirow{3}{*}{$\SLS_{-+}$} & \multirow{3}{*}{${\cal P}$DIII}
	& P & \Cb \\
	& & \Lr & \Rb \\
	& & \Li & \Rh \\ \hline
	\multirow{3}{*}{$\SLS_{-}$} & \multirow{3}{*}{${\cal P}$AII}
	& P & \Ca \\
	& & \Lr & \Rc \\
	& & \Li & \Rg \\ \hline
	\multirow{3}{*}{$\SLS_{-+}$} & \multirow{3}{*}{${\cal P}$CII}
	& P & \Cb \\
	& & \Lr & \Rd \\
	& & \Li & \Rf \\ \hline
	\multirow{2}{*}{$\SLS_{+}$} & \multirow{2}{*}{${\cal P}$C}
	& P & \Ca \\
	& & L & \Ree \\ \hline
	\multirow{3}{*}{$\SLS_{-+}$} & \multirow{3}{*}{${\cal P}$CI}
	& P & \Cb \\
	& & \Lr & \Rf \\
	& & \Li & \Rd \\ \hline \hline
    \multicolumn{12}{c}{continued on next page}
  \end{tabular}
\end{table}

\begin{table}[H]
	\centering
     \begin{tabular}{cccccccccccc} % \hline \hline
%    ~SLS~ & ~${\cal P}$AZ class~ & ~Gap~ & Classifying space & ~$p=0$~ & ~$p=1$~ & ~$p=2$~ & ~$p=3$~& ~$p=4$~ & ~$p=5$~ & ~$p=6$~ & ~$p=7$~ \\ \hline
    \multicolumn{12}{c}{TABLE \ref{tab: real AZ + SLS} --- continued} \\ \hline \hline
	\multirow{3}{*}{$\SLS_{--}$} & \multirow{3}{*}{${\cal P}$BDI}
	& P & \Rc \\
	& & \Lr & \Ca \\
	& & \Li & \Rcc \\ \hline
	\multirow{3}{*}{$\SLS_{++}$} & \multirow{3}{*}{${\cal P}$DIII}
	& P & \Ree \\
	& & \Lr & \Ca \\
	& & \Li & \Ca \\ \hline
	\multirow{3}{*}{$\SLS_{--}$} & \multirow{3}{*}{${\cal P}$CII}
	& P & \Rg \\
	& & \Lr & \Ca \\
	& & \Li & \Rgg \\ \hline
	\multirow{3}{*}{$\SLS_{++}$} & \multirow{3}{*}{${\cal P}$CI}
	& P & \Ra \\
	& & \Lr & \Ca \\
	& & \Li & \Ca \\ \hline \hline
	\multirow{3}{*}{$\SLS_{+}$} & \multirow{3}{*}{${\cal P}$AI}
	& P & \Raa \\
	& & \Lr & \Ra \\
	& & \Li & \Ra \\ \hline
	\multirow{3}{*}{$\SLS_{+-}$} & \multirow{3}{*}{${\cal P}$BDI}
	& P & \Rbb \\
	& & \Lr & \Rb \\
	& & \Li & \Rb \\ \hline
	\multirow{2}{*}{$\SLS_{-}$} & \multirow{2}{*}{${\cal P}$D}
	& P & \Rcc \\
	& & L & \Rc \\ \hline
	\multirow{3}{*}{$\SLS_{+-}$} & \multirow{3}{*}{${\cal P}$DIII}
	& P & \Rdd \\
	& & \Lr & \Rd \\
	& & \Li & \Rd \\ \hline
	\multirow{3}{*}{$\SLS_{+}$} & \multirow{3}{*}{${\cal P}$AII}
	& P & \Reee \\
	& & \Lr & \Ree \\
	& & \Li & \Ree \\ \hline
	\multirow{3}{*}{$\SLS_{+-}$} & \multirow{3}{*}{${\cal P}$CII}
	& P & \Rff \\
	& & \Lr & \Rf \\
	& & \Li & \Rf \\ \hline
	\multirow{2}{*}{$\SLS_{-}$} & \multirow{2}{*}{${\cal P}$C}
	& P & \Rgg \\
	& & L & \Rg \\ \hline
	\multirow{3}{*}{$\SLS_{+-}$} & \multirow{3}{*}{${\cal P}$CI}
	& P & \Rhh \\
	& & \Lr & \Rh \\
	& & \Li & \Rh \\ \hline \hline
  \end{tabular}
\end{table}
%%%%%%%%%%

%%%%% Complex AZ + pseudo-Hermiticity %%%%%
\begin{table}[H]
	\centering
	\caption{Topological classification table of exceptional points in complex AZ symmetry class with pseudo-Hermiticity (pH) $\eta$. Exceptional points are classified according to the symmetry class, the codimension $p$, and the definition of a complex-energy point (P) or line (L) gap. The subscript of $\eta$ specifies the commutation ($+$) or anticommutation ($-$) relation to chiral symmetry. The codimension $p$ is defined as $p := d-d_{\rm EP}$ with the spatial dimension $d$ and the dimension $d_{\rm EP}$ of the exceptional point. The subscript of L specifies the line gap for the real or imaginary part of the complex spectrum.\\}
		\label{tab: complex AZ + pH}
     \begin{tabular}{cccccccccccc} \hline \hline
    ~pH~ & ~AZ class~ & ~Gap~ & Classifying space & ~$p=0$~ & ~$p=1$~ & ~$p=2$~ & ~$p=3$~& ~$p=4$~ & ~$p=5$~ & ~$p=6$~ & ~$p=7$~ \\ \hline
	\multirow{3}{*}{$\eta$} & \multirow{3}{*}{A}
	& P & \Cb \\
	& & \Lr & \Cbb \\
	& & \Li & \Ca \\ \hline
	\multirow{3}{*}{$\eta_{+}$} & \multirow{3}{*}{AIII}
	& P & \Ca \\
	& & \Lr & \Caa \\
	& & \Li & \Caa \\ \hline \hline
	\multirow{3}{*}{$\eta_{-}$} & \multirow{3}{*}{AIII}
	& P & \Cbb \\
	& & \Lr & \Cb \\
	& & \Li & \Cb \\ \hline \hline
  \end{tabular}
\end{table}
%%%%%%%%%%

%%%%% Real AZ + pseudo-Hermiticity %%%%%
\begin{table}[H]
	\centering
	\caption{Topological classification table of exceptional points in real ${\cal P}$AZ symmetry class with pseudo-Hermiticity (pH) $\eta$. Exceptional points are classified according to the symmetry class, the codimension $p$, and the definition of a complex-energy point (P) or line (L) gap. The subscript of $\eta$ specifies the commutation ($+$) or anticommutation ($-$) relation to $\mathcal{P}\mathcal{T}_{+}$ symmetry and/or $\mathcal{C}_{-}\mathcal{P}$ symmetry; for the symmetry classes that involve both $\mathcal{P}\mathcal{T}_{+}$ and $\mathcal{C}_{-}\mathcal{P}$ symmetries (${\cal P}$BDI, ${\cal P}$DIII, ${\cal P}$CII, and ${\cal P}$CI), the first and second subscripts specify the relation to $\mathcal{P}\mathcal{T}_{+}$ and $\mathcal{C}_{-}\mathcal{P}$ symmetries, respectively. The codimension $p$ is defined as $p := d-d_{\rm EP}$ with the spatial dimension $d$ and the dimension $d_{\rm EP}$ of the exceptional point. The subscript of L specifies the line gap for the real or imaginary part of the complex spectrum.\\}
		\label{tab: real AZ + pH}
     \begin{tabular}{cccccccccccc} \hline \hline
    ~pH~ & ~${\cal P}$AZ class~ & ~Gap~ & Classifying space & ~$p=0$~ & ~$p=1$~ & ~$p=2$~ & ~$p=3$~& ~$p=4$~ & ~$p=5$~ & ~$p=6$~ & ~$p=7$~ \\ \hline
    	\multirow{3}{*}{$\eta_{+}$} & \multirow{3}{*}{${\cal P}$AI}
	& P & \Rh \\
	& & \Lr & \Rhh \\
	& & \Li & \Ra \\ \hline
	\multirow{3}{*}{$\eta_{++}$} & \multirow{3}{*}{${\cal P}$BDI}
	& P & \Ra \\
	& & \Lr & \Raa \\
	& & \Li & \Raa \\ \hline
	\multirow{3}{*}{$\eta_{+}$} & \multirow{3}{*}{${\cal P}$D}
	& P & \Rb \\
	& & \Lr & \Rbb \\
	& & \Li & \Ra \\ \hline
	\multirow{3}{*}{$\eta_{++}$} & \multirow{3}{*}{${\cal P}$DIII}
	& P & \Rc \\
	& & \Lr & \Rcc \\
	& & \Li & \Ca \\ \hline
	\multirow{3}{*}{$\eta_{+}$} & \multirow{3}{*}{${\cal P}$AII}
	& P & \Rd \\
	& & \Lr & \Rdd \\
	& & \Li & \Ree \\ \hline
	\multirow{3}{*}{$\eta_{++}$} & \multirow{3}{*}{${\cal P}$CII}
	& P & \Ree \\
	& & \Lr & \Reee \\
	& & \Li & \Reee \\ \hline
	\multirow{3}{*}{$\eta_{+}$} & \multirow{3}{*}{${\cal P}$C}
	& P & \Rf \\
	& & \Lr & \Rff \\
	& & \Li & \Ree \\ \hline
	\multirow{3}{*}{$\eta_{++}$} & \multirow{3}{*}{${\cal P}$CI}
	& P & \Rg \\
	& & \Lr & \Rgg \\
	& & \Li & \Ca \\ \hline \hline
    \multicolumn{12}{c}{continued on next page}
  \end{tabular}
\end{table}

\begin{table}[H]
	\centering
     \begin{tabular}{cccccccccccc} %\hline \hline
%    ~SLS~ & ~AZ class~ & ~Gap~ & Classifying space & ~$p=0$~ & ~$p=1$~ & ~$p=2$~ & ~$p=3$~ \\ \hline
    \multicolumn{12}{c}{TABLE \ref{tab: real AZ + pH} --- continued} \\ \hline \hline
	\multirow{3}{*}{$\eta_{+-}$} & \multirow{3}{*}{${\cal P}$BDI}
	& P & \Cb \\
	& & \Lr & \Rh \\
	& & \Li & \Rb \\ \hline
	\multirow{3}{*}{$\eta_{-+}$} & \multirow{3}{*}{${\cal P}$DIII}
	& P & \Cb \\
	& & \Lr & \Rb \\
	& & \Li & \Rh \\ \hline
	\multirow{3}{*}{$\eta_{+-}$} & \multirow{3}{*}{${\cal P}$CII}
	& P & \Cb \\
	& & \Lr & \Rd \\
	& & \Li & \Rf \\ \hline
	\multirow{3}{*}{$\eta_{-+}$} & \multirow{3}{*}{${\cal P}$CI}
	& P & \Cb \\
	& & \Lr & \Rf \\
	& & \Li & \Rd \\ \hline \hline
    	\multirow{3}{*}{$\eta_{-}$} & \multirow{3}{*}{${\cal P}$AI}
	& P & \Rb \\
	& & \Lr & \Cb \\
	& & \Li & \Rc \\ \hline
	\multirow{3}{*}{$\eta_{--}$} & \multirow{3}{*}{${\cal P}$BDI}
	& P & \Rc \\
	& & \Lr & \Ca \\
	& & \Li & ~\Rcc~ \\ \hline
	\multirow{3}{*}{$\eta_{-}$} & \multirow{3}{*}{${\cal P}$D}
	& P & \Rd \\
	& & \Lr & \Cb \\
	& & \Li & \Rc \\ \hline
	\multirow{3}{*}{$\eta_{--}$} & \multirow{3}{*}{${\cal P}$DIII}
	& P & \Ree \\
	& & \Lr & \Ca \\
	& & \Li & \Ca \\ \hline
	\multirow{3}{*}{$\eta_{-}$} & \multirow{3}{*}{${\cal P}$AII}
	& P & \Rf \\
	& & \Lr & \Cb \\
	& & \Li & \Rg \\ \hline
	\multirow{3}{*}{$\eta_{--}$} & \multirow{3}{*}{${\cal P}$CII}
	& P & \Rg \\
	& & \Lr & \Ca \\
	& & \Li & \Rgg \\ \hline
	\multirow{3}{*}{$\eta_{-}$} & \multirow{3}{*}{${\cal P}$C}
	& P & \Rh \\
	& & \Lr & \Cb \\
	& & \Li & \Rg \\ \hline
	\multirow{3}{*}{$\eta_{--}$} & \multirow{3}{*}{${\cal P}$CI}
	& P & \Ra \\
	& & \Lr & \Ca \\
	& & \Li & \Ca \\ \hline \hline
	\multirow{3}{*}{$\eta_{-+}$} & \multirow{3}{*}{${\cal P}$BDI}
	& P & \Rbb \\
	& & \Lr & \Rb \\
	& & \Li & \Rb \\ \hline
	\multirow{3}{*}{$\eta_{+-}$} & \multirow{3}{*}{${\cal P}$DIII}
	& P & \Rdd \\
	& & \Lr & \Rd \\
	& & \Li & \Rd \\ \hline
	\multirow{3}{*}{$\eta_{-+}$} & \multirow{3}{*}{${\cal P}$CII}
	& P & \Rff \\
	& & \Lr & \Rf \\
	& & \Li & \Rf \\ \hline
	\multirow{3}{*}{$\eta_{+-}$} & \multirow{3}{*}{${\cal P}$CI}
	& P & \Rhh \\
	& & \Lr & \Rh \\
	& & \Li & \Rh \\ \hline \hline
  \end{tabular}
\end{table}
%%%%%%%%%%

\clearpage
%%%%%%%%%%
\section{SIII.~Derivations of topological classification}

Topology of exceptional points is determined in a similar manner to that of Fermi surfaces in Hermitian systems~\cite{Morimoto-14, Kobayashi-14, Zhao-16}. In fact, whereas a complex-energy gap is closed at the gapless point, it may be open and a topological invariant can be assigned on a $\left( p-1 \right)$-dimensional surface $S^{p-1}$ that encloses the gapless point. On this enclosing surface $S^{p-1}$, one of the kinetic terms in Eq.~(\ref{eq: Hamiltonian; near-gapless}) in the main text can be regarded as a mass term and the topology of the exceptional point is determined by classifying this mass term, which reduces to the extension of Clifford algebra~\cite{Schnyder-Ryu-review, Karoubi} as with the gapped case.

As discussed in the main text, non-Hermitian topology crucially depends on the types of complex-energy gaps~\cite{KSUS-18}. In particular, a proper representation of non-Hermitian Dirac matrices depends on the types of complex-energy gaps. We first consider the case where the Hamiltonian has a point gap on $S^{p-1}$. Without loss of generality, we take the reference point of the gap as $E=0$. In the presence of the point gap [i.e., $\forall\,{\bm k}~~E \left( {\bm k} \right) \neq 0$], a generic non-Hermitian Hamiltonian can be flattened into a unitary matrix~\cite{KSUS-18}. As a result, a set of non-Hermitian Dirac matrices $\{ \gamma_{i}^{\rm P} \}$ for the point gap does not necessarily follow Clifford algebra. Instead, a set of Hermitian matrices $\{ \gamma_{i} \}$ defined by 
\begin{equation}
\gamma_{i} := \begin{pmatrix}
0 & \gamma_{i}^{\rm P} \\ (\gamma_{i}^{\rm P})^{\dag} & 0
\end{pmatrix},\quad
\gamma_{i}^{\dag} = \gamma_{i}
	\label{SM: eq - Hermitianization}
\end{equation}
follows Clifford algebra
\begin{equation}
\{ \gamma_{i}, \gamma_{j} \} = 2 \delta_{ij} 
~~\Leftrightarrow~~
\gamma_{i}^{\rm P} (\gamma_{j}^{\rm P})^{\dag} + \gamma_{j}^{\rm P} (\gamma_{i}^{\rm P})^{\dag} = 2\delta_{ij}.
\end{equation} 
Accordingly, symmetries impose the following constraints on the extended Hermitian Dirac matrices:
\begin{eqnarray}
({\sf P}{\sf T}_{+})\,\gamma_{i}^{*}\,({\sf P}{\sf T}_{+})^{-1}
&=& \gamma_{i},\quad
{\sf P}{\sf T}_{+} := \left( \begin{array}{@{\,}cc@{\,}} 
	{\cal P}{\cal T}_{+} & 0 \\
	0 & {\cal P}{\cal T}_{+} \\ 
	\end{array} \right); \label{SM: eq - PT} \\
({\sf C}_{+} {\sf P})\,\gamma_{i}^{*}\,({\sf C}_{+} {\sf P})^{-1}
&=& \gamma_{i},\quad
{\sf C}_{+} {\sf P} := \left( \begin{array}{@{\,}cc@{\,}} 
	0 & {\cal C}_{+} {\cal P} \\
	{\cal C}_{+} {\cal P} & 0\\ 
	\end{array} \right); \\
({\sf C}_{-} {\sf P})\,\gamma_{i}^{*}\,({\sf C}_{-} {\sf P})^{-1}
&=& - \gamma_{i},\quad
{\sf C}_{-} {\sf P} := \left( \begin{array}{@{\,}cc@{\,}} 
	0 & {\cal C}_{-} {\cal P} \\
	{\cal C}_{-} {\cal P} & 0 \\ 
	\end{array} \right); \label{SM: eq - CP} \\
({\sf P}{\sf T}_{-})\,\gamma_{i}^{*}\,({\sf P}{\sf T}_{-})^{-1}
&=& - \gamma_{i},\quad
{\sf P}{\sf T}_{-} := \left( \begin{array}{@{\,}cc@{\,}} 
	{\cal P}{\cal T}_{-} & 0 \\
	0 & {\cal P}{\cal T}_{-} \\ 
	\end{array} \right); \\
{\sf \Gamma}\,\gamma_{i}\,{\sf \Gamma}^{-1}
&=& - \gamma_{i},\quad
{\sf \Gamma} := \left( \begin{array}{@{\,}cc@{\,}} 
	0 & {\Gamma} \\
	{\Gamma} & 0 \\ 
	\end{array} \right); \label{SM: eq - chiral} \\
{\sf S}\,\gamma_{i} {\sf S}^{-1}
&=& - \gamma_{i},\quad
{\sf S} := \left( \begin{array}{@{\,}cc@{\,}} 
	{\cal S} & 0 \\
	0 & {\cal S} \\ 
	\end{array} \right).
\end{eqnarray}
Moreover, $\gamma_{i}$ respects additional chiral (sublattice) symmetry by construction:
\begin{equation}
{\sf \Sigma}\,\gamma_{i} {\sf \Sigma}^{-1}
= - \gamma_{i},\quad
{\sf \Sigma} := \left( \begin{array}{@{\,}cc@{\,}} 
	1 & 0 \\
	0 & -1 \\ 
	\end{array} \right).
	\label{SM: eq - sigma}
\end{equation}
The Dirac matrices $\gamma_{i}$'s and the above symmetry operators form Clifford algebra, and the topology of the ``point-gapless" point is determined by the extension of Clifford algebra~\cite{Schnyder-Ryu-review}.
 
On the other hand, a non-Hermitian Hamiltonian with a line gap can be flattened into a Hermitian (an anti-Hermitian) matrix in the presence of a real (imaginary) gap [i.e., $\forall\,{\bm k}~~{\rm Re}\,E \left( {\bm k} \right) \neq 0$ (${\rm Im}\,E \left( {\bm k} \right) \neq 0$)]~\cite{KSUS-18}. Without loss of generality, the reference line of the gap is taken to be ${\rm Re}\,E = 0$ (${\rm Im}\,E = 0$). As a result, a set of Dirac matrices $\{ \gamma_{i}^{\rm L} \}$ follows Clifford algebra. As with the case of the point gap, the Dirac matrices $\gamma_{i}^{\rm L}$'s and the symmetry operators form Clifford algebra and the topology of the ``line-gapless" point is determined by its extension. We note that the following formulas are useful for the extension of Clifford algebra~\cite{Karoubi}:
\begin{equation}
Cl_{p} \rightarrow Cl_{p+1}~\Leftrightarrow~{\cal C}_{p},\quad 
Cl_{p, q} \rightarrow Cl_{p, q+1}~\Leftrightarrow~{\cal R}_{q-p},\quad 
Cl_{p, q} \rightarrow Cl_{p+1, q}~\Leftrightarrow~{\cal R}_{p+2-q},
\end{equation}
where $Cl_{n}$ and $Cl_{m, n}$ are complex and real Clifford algebras, respectively, and ${\cal C}_{n}$ and ${\cal R}_{n}$ are the corresponding classifying spaces, which have Bott periodicity, i.e., ${\cal C}_{n+2} = {\cal C}_{n}$, ${\cal R}_{n+8} = {\cal R}_{n}$. The classifying spaces and their homotopy groups are summarized in Table~\ref{tab: classifying space - homotopy}.

\begin{table}[H]
	\centering
	\caption{Classifying spaces and their homotopy groups.}
     \begin{tabular}{ccc} \hline \hline
     & ~~Classifying space~~~ & ~~$\pi_{0} \left( \star \right)$~~ \\ \hline
     ~~${\cal C}_{0}$~~ & $\left[ {\rm U} \left( m+n\right) / {\rm U} \left( m \right) \times {\rm U} \left( n \right) \right] \times \mathbb{Z}$ & $\mathbb{Z}$ \\
     ${\cal C}_{1}$ & ${\rm U} \left( n\right)$ & $0$ \\ \hline
     ${\cal R}_{0}$ & $\left[ {\rm O} \left( m+n\right) / {\rm O} \left( m \right) \times {\rm O} \left( n \right) \right] \times \mathbb{Z}$ & $\mathbb{Z}$ \\
     ${\cal R}_{1}$ & ${\rm O} \left( n\right)$ & $\mathbb{Z}_{2}$ \\
     ${\cal R}_{2}$ & ${\rm O} \left( 2n \right) / {\rm U} \left( n \right)$ & $\mathbb{Z}_{2}$ \\
     ${\cal R}_{3}$ & ${\rm U} \left( 2n \right) / {\rm Sp} \left( n \right)$ & $0$ \\
     ${\cal R}_{4}$ & ~~~$\left[ {\rm Sp} \left( m+n\right) / {\rm Sp} \left( m \right) \times {\rm Sp} \left( n \right) \right] \times \mathbb{Z}$~~~ & $\mathbb{Z}$ \\
     ${\cal R}_{5}$ & ${\rm Sp} \left( n\right)$ & $0$ \\
     ${\cal R}_{6}$ & ${\rm Sp} \left( n \right) / {\rm U} \left( n \right)$ & $0$ \\
     ~~${\cal R}_{7}$~~ & ${\rm U} \left( n \right) / {\rm O} \left( n \right)$ & $0$ \\ \hline \hline
     \end{tabular}
  	\label{tab: classifying space - homotopy}
\end{table}

In Table~\ref{tab: extension}, we explicitly provide the extensions of Clifford algebras and the corresponding classifying spaces for all the 38 symmetry classes and two types of complex-energy gaps, which leads to classification tables~\ref{tab: complex + real AZ}-\ref{tab: real AZ + pH} including Table~\ref{tab: classification} in the main text. We respectively denote $\mathcal{P}\mathcal{T}_{+}$ (${\sf P}{\sf T}_{+}$) and $\mathcal{C}_{-} \mathcal{P}$ (${\sf C}_{-}{\sf P}$) symmetries as $\mathcal{PT}$ (${\sf PT}$) and $\mathcal{CP}$ (${\sf CP}$) symmetries for simplicity in Table~\ref{tab: extension}. 

Below we explicitly provide a couple of examples of the classification procedure. First, let us consider class AIII. A non-Hermitian Hamiltonian in class AIII respects chiral symmetry [Eq.~(\ref{eq: chiral symmetry}) in the main text]: 
\begin{equation}
\Gamma H^{\dag} \left( {\bm k} \right) \Gamma^{-1} = - H \left( {\bm k} \right).
\end{equation} 
In the presence of a point gap, the extended Hermitian Dirac matrices $\gamma_{i}$'s defined as Eq.~(\ref{SM: eq - Hermitianization}) respect chiral symmetries defined by Eqs.~(\ref{SM: eq - chiral}) and (\ref{SM: eq - sigma}). Since ${\sf \Sigma}$ and ${\sf \Gamma}$ anticommute with each other, a set of operators
\begin{equation}
\{ \gamma_{1}, \cdots, \gamma_{p}, {\sf \Sigma}, {\sf \Gamma} \}
\end{equation}
forms complex Clifford algebra $Cl_{p+2}$. Hence, when one of the Dirac matrices is regarded as a mass term, we have an extension of Clifford algebra $Cl_{p+1} \rightarrow Cl_{p+2}$, which leads to the classifying space ${\cal C}_{p+1}$. The topology of point-gapless points with codimension $p$ in class AIII is determined by $\pi_{0} \left( {\cal C}_{p+1} \right)$ [see Table~\ref{tab: classifying space - homotopy} for $\pi_{0} \left( {\cal C}_{p+1} \right)$]. By contrast, different classifying spaces characterize exceptional points when a line gap is relevant. In the presence of a real gap, $\gamma_{i}^{\rm L} = \gamma_{i}$ respects chiral symmetry, and hence a set of operators  
\begin{equation}
\{ \gamma_{1}, \cdots, \gamma_{p}, \Gamma \}
\end{equation}
forms complex Clifford algebra $Cl_{p+1}$, which leads to the classifying space ${\cal C}_{p}$ with its topology $\pi_{0} \left( {\cal C}_{p} \right)$. On the other hand, in the presence of an imaginary gap, $\gamma_{i}^{\rm L} = \ii \gamma_{i}$ respects chiral symmetry, i.e.,
\begin{equation}
\Gamma\,\gamma_{i}\,\Gamma^{-1} = \gamma_{i}.
\end{equation}
Hence, a set of operators 
\begin{equation}
\{ \gamma_{1}, \cdots, \gamma_{p} \}
\end{equation}
forms complex Clifford algebra $Cl_{p}$ and all of them commute with $\Gamma$, which leads to the classifying space ${\cal C}_{p-1} \times {\cal C}_{p-1}$ with its topology $\pi_{0} \left( {\cal C}_{p-1} \right) \oplus \pi_{0} \left( {\cal C}_{p-1} \right)$.

Let us also consider class ${\cal P}$CI as another example. A non-Hermitian Hamiltonian in class ${\cal P}$CI respects $\PT$ symmetry and $\CP$ symmetry [Eqs.~(\ref{eq: PT symmetry}) and (\ref{eq: CP symmetry}) in the main text]:
\begin{eqnarray}
(\PT)\,H^{*} \left( {\bm k} \right) (\PT)^{-1} = H \left( {\bm k} \right),\quad
(\PT) (\PT)^{*} = +1; \\
(\CP)\,H^{T} \left( {\bm k} \right) (\CP)^{-1} = - H \left( {\bm k} \right),\quad
(\CP) (\CP)^{*} = -1.
\end{eqnarray}
In the presence of a point gap, the extended Hermitian Dirac matrices $\gamma_{i}$'s defined as Eq.~(\ref{SM: eq - Hermitianization}) respect Eqs.~(\ref{SM: eq - PT}), (\ref{SM: eq - CP}), and (\ref{SM: eq - sigma}). Since the symmetry operators satisfy
 \begin{equation}
 [ {\sf PT}\mathcal{K}, {\sf CP}\mathcal{K} ] = [ {\sf PT}\mathcal{K}, {\sf \Sigma} ] = \{ {\sf CP}\mathcal{K}, {\sf \Sigma} \} = 0,
 \end{equation}
a set of operators
\begin{equation}
\{ {\sf CP}{\cal K}, J{\sf CP}{\cal K}; \gamma_{1}, \cdots, \gamma_{p}, {\sf \Sigma}, J ({\sf CP}{\cal K}) ({\sf PT} {\cal K}) \}
\end{equation}
forms real Clifford algebra $Cl_{2, p+2}$ with the imaginary unit $J$ and complex conjugation ${\cal K}$. Hence, when one of the Dirac matrices is regarded as a mass term, we have an extension of Clifford algebra $Cl_{2, p+1} \rightarrow Cl_{2, p+2}$, which leads to the classifying space ${\cal R}_{p+7}$. The topology of point-gapless points with codimension $p$ in class ${\cal P}$CI is determined by $\pi_{0} \left( {\cal R}_{p+7} \right)$. By contrast, different classifying spaces characterize exceptional points when a line gap is relevant. In the presence of a real gap, $\gamma_{i}^{\rm L} = \gamma_{i}$ respects $\PT$ and $\CP$ symmetries, and hence a set of operators  
\begin{equation}
\{ {\cal CP}{\cal K}, J{\cal CP}{\cal K}; \gamma_{1}, \cdots, \gamma_{p}, J ({\cal CP}{\cal K}) ({\cal PT} {\cal K}) \}
\end{equation}
forms real Clifford algebra $Cl_{2, p+1}$, which leads to the classifying space ${\cal R}_{p+6}$ with its topology $\pi_{0} \left( {\cal R}_{p+6} \right)$. On the other hand, in the presence of an imaginary gap, $\gamma_{i}^{\rm L} = \ii \gamma_{i}$ respects $\PT$ and $\CP$ symmetries, i.e.,
\begin{equation}
(\PT)\,\gamma_{i}^{*}\,(\PT)^{-1} = -\gamma_{i},\quad
(\CP)\,\gamma_{i}^{*}\,(\CP)^{-1} = -\gamma_{i}.
\end{equation}
Hence, a set of operators 
\begin{equation}
\{ {\cal CPK}, J{\cal CPK}; \gamma_{1}, \cdots, \gamma_{p} \}
\end{equation}
forms real Clifford algebra $Cl_{2, p}$ and all of them commute with $({\cal CPK}) ({\cal PTK})$. Noting $[({\cal CPK}) ({\cal PTK})]^{2} = -1$, we have the classifying space ${\cal C}_{p+1}$ with its topology $\pi_{0} \left( {\cal C}_{p+1} \right)$. The extensions of Clifford algebra and the corresponding classifying spaces for all the other symmetry classes are similarly determined as Table~\ref{tab: extension}.

%\clearpage
%%%%%%%%%%
\begin{table}[H]
	\centering
	\footnotesize
	\caption{Clifford algebra extensions and classifying spaces. $\mathcal{P}\mathcal{T}_{+}$ (${\sf P}{\sf T}_{+}$) and $\mathcal{C}_{-}\mathcal{P}$ (${\sf C}_{-}{\sf P}$) symmetries are respectively denoted as $\mathcal{PT}$ (${\sf PT}$) and $\mathcal{CP}$ (${\sf CP}$) symmetries  for simplicity.\\}
		\label{tab: extension}
     \begin{tabular}{ccccc} \hline \hline
    ~Symmetry class~ & ~Gap~ & ~Generator~ & ~Extension~ & Classifying space \\ \hline
	\multirow{2}{*}{A} 
	& P & $\{ \gamma_{1}, \cdots, \gamma_{p}, {\sf \Sigma} \}$ & $Cl_{p} \rightarrow Cl_{p+1}$ & ${\cal C}_{p}$ \\ 
	& L & $\{ \gamma_{1}, \cdots, \gamma_{p} \}$ & $Cl_{p-1} \rightarrow Cl_{p}$ & ${\cal C}_{p+1}$ \\ \hline
	\multirow{3}{*}{AIII} 
	& P & $\{ \gamma_{1}, \cdots, \gamma_{p}, {\sf \Sigma}, {\sf \Gamma} \}$ & $Cl_{p+1} \rightarrow Cl_{p+2}$ & ${\cal C}_{p+1}$ \\ 
	& \Lr & $\{ \gamma_{1}, \cdots, \gamma_{p}, \Gamma \}$ & $Cl_{p} \rightarrow Cl_{p+1}$ & ${\cal C}_{p}$ \\
	& \Li & $\{ \gamma_{1}, \cdots, \gamma_{p} \} \otimes \{ \Gamma \}$ & $Cl_{p-1} \times Cl_{p-1} \rightarrow Cl_{p} \times Cl_{p}$ & ${\cal C}_{p+1} \times {\cal C}_{p+1}$ \\ \hline \hline % complex AZ
	\multirow{3}{*}{${\cal P}$AI} 
	& P & $\{ J\gamma_{1}, \cdots, J\gamma_{p}, J{\sf \Sigma}; {\sf PT} {\cal K}, J {\sf PT} {\cal K} \}$ & $Cl_{p, 2} \rightarrow Cl_{p+1, 2}$ & ${\cal R}_{p}$ \\ 
	& \Lr & $\{ J\gamma_{1}, \cdots, J\gamma_{p}; {\cal PTK}, J{\cal PTK} \}$ & $Cl_{p-1, 2} \rightarrow Cl_{p, 2}$ & ${\cal R}_{p+7}$ \\
	& \Li & $\{ \gamma_{1}, \cdots, \gamma_{p}, {\cal PTK}, J{\cal PTK}\}$ & $Cl_{0, p+1} \rightarrow Cl_{0, p+2}$ & ${\cal R}_{p+1}$ \\ \hline
	\multirow{3}{*}{${\cal P}$BDI} 
	& P & ~$\{ J ({\sf CP}{\cal K}) ({\sf PT} {\cal K}); \gamma_{1}, \cdots, \gamma_{p}, {\sf \Sigma}, {\sf CP}{\cal K}, J{\sf CP}{\cal K} \}$~ & $Cl_{1, p+2} \rightarrow Cl_{1, p+3}$ & ${\cal R}_{p+1}$ \\ 
	& \Lr & $\{ J ({\cal CP}{\cal K}) ({\cal PT} {\cal K}); \gamma_{1}, \cdots, \gamma_{p}, {\cal CP}{\cal K}, J{\cal CP}{\cal K} \}$ & $Cl_{1, p+1} \rightarrow Cl_{1, p+2}$ & ${\cal R}_{p}$ \\
	& \Li & $\{ \gamma_{1}, \cdots, \gamma_{p}, {\cal CPK}, J{\cal CPK} \} \otimes \{ ({\cal CPK}) ({\cal PTK}) \}$ & $Cl_{0, p+1} \times Cl_{0, p+1} \rightarrow Cl_{0, p+2} \times Cl_{0, p+2}$ & ${\cal R}_{p+1} \times {\cal R}_{p+1}$ \\ \hline
	\multirow{2}{*}{${\cal P}$D} 
	& P & $\{ \gamma_{1}, \cdots, \gamma_{p}, {\sf \Sigma}, {\sf CP} {\cal K}, J {\sf CP} {\cal K} \}$ & $Cl_{0, p+2} \rightarrow Cl_{0, p+3}$ & ${\cal R}_{p+2}$ \\ 
	& L & $\{ \gamma_{1}, \cdots, \gamma_{p}, {\cal CP} {\cal K}, J {\cal CP} {\cal K} \}$ & $Cl_{0, p+1} \rightarrow Cl_{0, p+2}$ & ${\cal R}_{p+1}$ \\ \hline
	\multirow{3}{*}{${\cal P}$DIII} 
	& P & ~$\{ \gamma_{1}, \cdots, \gamma_{p}, {\sf \Sigma}, {\sf CP}{\cal K}, J{\sf CP}{\cal K}, J ({\sf CP}{\cal K}) ({\sf PT} {\cal K}) \}$~ & $Cl_{0, p+3} \rightarrow Cl_{0, p+4}$ & ${\cal R}_{p+3}$ \\ 
	& \Lr & $\{ \gamma_{1}, \cdots, \gamma_{p}, {\cal CP}{\cal K}, J{\cal CP}{\cal K}, J ({\cal CP}{\cal K}) ({\cal PT} {\cal K}) \}$ & $Cl_{0, p+2} \rightarrow Cl_{0, p+3}$ & ${\cal R}_{p+2}$ \\
	& \Li & $\{ \gamma_{1}, \cdots, \gamma_{p}, {\cal CPK}, J{\cal CPK} \} \otimes \{ ({\cal CPK}) ({\cal PTK}) \}$ & $Cl_{p+1} \rightarrow Cl_{p+2}$ & ${\cal C}_{p+1}$ \\ \hline
	\multirow{3}{*}{${\cal P}$AII} 
	& P & $\{ J\gamma_{1}, \cdots, J\gamma_{p}, J{\sf \Sigma}, {\sf PT} {\cal K}, J {\sf PT} {\cal K} \}$ & $Cl_{p+2, 0} \rightarrow Cl_{p+3, 0}$ & ${\cal R}_{p+4}$ \\ 
	& \Lr & $\{ J\gamma_{1}, \cdots, J\gamma_{p}, {\cal PTK}, J{\cal PTK} \}$ & $Cl_{p+1, 0} \rightarrow Cl_{p+2, 0}$ & ${\cal R}_{p+3}$ \\
	& \Li & $\{ {\cal PTK}, J{\cal PTK}; \gamma_{1}, \cdots, \gamma_{p}\}$ & $Cl_{2, p-1} \rightarrow Cl_{2, p}$ & ${\cal R}_{p+5}$ \\ \hline
	\multirow{3}{*}{${\cal P}$CII} 
	& P & ~$\{ {\sf CP}{\cal K}, J{\sf CP}{\cal K}, J ({\sf CP}{\cal K}) ({\sf PT} {\cal K}); \gamma_{1}, \cdots, \gamma_{p}, {\sf \Sigma} \}$~ & $Cl_{3, p} \rightarrow Cl_{3, p+1}$ & ${\cal R}_{p+5}$ \\ 
	& \Lr & ~$\{ {\cal CP}{\cal K}, J{\cal CP}{\cal K}, J ({\cal CP}{\cal K}) ({\cal PT} {\cal K}); \gamma_{1}, \cdots, \gamma_{p} \}$~ & $Cl_{3, p-1} \rightarrow Cl_{3, p}$ & ${\cal R}_{p+4}$ \\
	& \Li & $\{ {\cal CPK}, J{\cal CPK}; \gamma_{1}, \cdots, \gamma_{p} \} \otimes \{ ({\cal CPK}) ({\cal PTK}) \}$ & $Cl_{2, p-1} \times Cl_{2, p-1} \rightarrow Cl_{2, p} \times Cl_{2, p}$ & ${\cal R}_{p+5} \times {\cal R}_{p+5}$ \\ \hline
	\multirow{2}{*}{${\cal P}$C} 
	& P & $\{ {\sf CP} {\cal K}, J {\sf CP} {\cal K}; \gamma_{1}, \cdots, \gamma_{p}, {\sf \Sigma} \}$ & $Cl_{2, p} \rightarrow Cl_{2, p+1}$ & ${\cal R}_{p+6}$ \\ 
	& L & $\{ {\cal CP} {\cal K}, J {\cal CP} {\cal K}; \gamma_{1}, \cdots, \gamma_{p} \}$ & $Cl_{2, p-1} \rightarrow Cl_{2, p}$ & ${\cal R}_{p+5}$ \\ \hline
	\multirow{3}{*}{${\cal P}$CI} 
	& P & ~$\{ {\sf CP}{\cal K}, J{\sf CP}{\cal K}; \gamma_{1}, \cdots, \gamma_{p}, {\sf \Sigma}, J ({\sf CP}{\cal K}) ({\sf PT} {\cal K}) \}$~ & $Cl_{2, p+1} \rightarrow Cl_{2, p+2}$ & ${\cal R}_{p+7}$ \\ 
	& \Lr & ~$\{ {\cal CP}{\cal K}, J{\cal CP}{\cal K}; \gamma_{1}, \cdots, \gamma_{p}, J ({\cal CP}{\cal K}) ({\cal PT} {\cal K}) \}$~ & $Cl_{2, p} \rightarrow Cl_{2, p+1}$ & ${\cal R}_{p+6}$ \\
	& \Li & $\{ {\cal CPK}, J{\cal CPK}; \gamma_{1}, \cdots, \gamma_{p} \} \otimes \{ ({\cal CPK}) ({\cal PTK}) \}$ & $Cl_{p+1} \rightarrow Cl_{p+2}$ & ${\cal C}_{p+1}$ \\ \hline \hline % real AZ
   \multicolumn{5}{c}{continued on next page}
  \end{tabular}
\end{table}

\begin{table}[H]
	\centering
	\scriptsize
     \begin{tabular}{ccccc} % \hline \hline
%    ~Symmetry class~ & ~Gap~ & ~Generator~ & ~Extension~ & Classifying space \\ \hline
    \multicolumn{5}{c}{TABLE \ref{tab: extension} --- continued} \\ \hline \hline
%    ~Symmetry class~ & ~Gap~ & ~Generator~ & ~Extension~ & Classifying space \\ \hline
	\multirow{2}{*}{${\cal P}\text{AI}^{\dag}$} 
	& P & $\{ J\gamma_{1}, \cdots, J\gamma_{p}; {\sf \Sigma}, {\sf C}_{+}{\sf P} {\cal K}, J {\sf C}_{+}{\sf P} {\cal K} \}$ & $Cl_{p-1, 3} \rightarrow Cl_{p, 3}$ & ${\cal R}_{p+6}$ \\ 
	& L & $\{ J\gamma_{1}, \cdots, J\gamma_{p}; {\cal C}_{+}{\cal P}{\cal K}, J{\cal C}_{+}{\cal P}{\cal K} \}$ & $Cl_{p-1, 2} \rightarrow Cl_{p, 2}$ & ${\cal R}_{p+7}$ \\ \hline
	\multirow{3}{*}{${\cal P}\text{BDI}^{\dag}$} 
	& P & ~$\{ {\sf \Sigma} ({\sf C}_{+}{\sf P} {\cal K}), J{\sf \Sigma} ({\sf C}_{+}{\sf P} {\cal K}); \gamma_{1}, \cdots, \gamma_{p}, {\sf \Sigma}, ({\sf P}{\sf T}_{-}{\cal K}) ({\sf C}_{+}{\sf P} {\cal K}) \}$~ & $Cl_{2, p+1} \rightarrow Cl_{2, p+2}$ & ${\cal R}_{p+7}$ \\ 
	& \Lr & $\{ J ({\cal P}{\cal T}_{-}{\cal K}) ({\cal C}_{+}{\cal P} {\cal K}); \gamma_{1}, \cdots, \gamma_{p}, {\cal P}{\cal T}_{-}{\cal K}, J{\cal P}{\cal T}_{-}{\cal K} \}$ & $Cl_{1, p+1} \rightarrow Cl_{1, p+2}$ & ${\cal R}_{p}$ \\
	& \Li & $\{ J\gamma_{1}, \cdots, J\gamma_{p}; {\cal C}_{+}{\cal P} {\cal K}, J {\cal C}_{+}{\cal P} {\cal K} \} \otimes \{ ({\cal P}{\cal T}_{-}{\cal K}) ({\cal C}_{+}{\cal P} {\cal K}) \}$ & $Cl_{p-1, 2} \times Cl_{p-1, 2} \rightarrow Cl_{p, 2} \times Cl_{p, 2}$ & ${\cal R}_{p+7} \times {\cal R}_{p+7}$ \\ \hline
	\multirow{3}{*}{${\cal P}\text{D}^{\dag}$} 
	& P & $\{ J{\sf \Sigma}; \gamma_{1}, \cdots, \gamma_{p}, {\sf P}{\sf T}_{-}{\cal K}, J{\sf P}{\sf T}_{-}{\cal K} \}$ & $Cl_{1, p+1} \rightarrow Cl_{1, p+2}$ & ${\cal R}_{p}$ \\ 
	& \Lr & $\{ \gamma_{1}, \cdots, \gamma_{p}, {\cal P}{\cal T}_{-}{\cal K}, J{\cal P}{\cal T}_{-}{\cal K} \}$ & $Cl_{0, p+1} \rightarrow Cl_{0, p+2}$ & ${\cal R}_{p+1}$ \\
	& \Li & $\{ J\gamma_{1}, \cdots, J \gamma_{p}; {\cal P}{\cal T}_{-}{\cal K}, J {\cal P}{\cal T}_{-}{\cal K} \}$ & $Cl_{p-1, 2} \rightarrow Cl_{p, 2}$ & ${\cal R}_{p+7}$ \\ \hline
	\multirow{3}{*}{${\cal P}\text{DIII}^{\dag}$} 
	& P & ~$\{ ({\sf P}{\sf T}_{-}{\cal K}) ({\sf C}_{+}{\sf P} {\cal K}); \gamma_{1}, \cdots, \gamma_{p}, {\sf \Sigma}, {\sf \Sigma} ({\sf C}_{+}{\sf P} {\cal K}), J{\sf \Sigma} ({\sf C}_{+}{\sf P} {\cal K}) \}$~ & $Cl_{1, p+2} \rightarrow Cl_{1, p+3}$ & ${\cal R}_{p+1}$ \\ 
	& \Lr & $\{ \gamma_{1}, \cdots, \gamma_{p}, {\cal P}{\cal T}_{-}{\cal K}, J {\cal P}{\cal T}_{-}{\cal K}, J ({\cal P}{\cal T}_{-}{\cal K}) ({\cal C}_{+}{\cal P} {\cal K}) \}$ & $Cl_{0, p+2} \rightarrow Cl_{0, p+3}$ & ${\cal R}_{p+2}$ \\
	& \Li & $\{ J\gamma_{1}, \cdots, J\gamma_{p}, {\cal C}_{+}{\cal P} {\cal K}, J {\cal C}_{+}{\cal P} {\cal K} \} \otimes \{ ({\cal P}{\cal T}_{-}{\cal K}) ({\cal C}_{+}{\cal P} {\cal K}) \}$ & $Cl_{p+1} \rightarrow Cl_{p+2}$ & ${\cal C}_{p+1}$ \\ \hline
	\multirow{2}{*}{${\cal P}\text{AII}^{\dag}$} 
	& P & $\{ J\gamma_{1}, \cdots, J\gamma_{p}, {\sf C}_{+}{\sf P} {\cal K}, J {\sf C}_{+}{\sf P} {\cal K}; {\sf \Sigma} \}$ & $Cl_{p+1, 1} \rightarrow Cl_{p+2, 1}$ & ${\cal R}_{p+2}$ \\ 
	& L & $\{ J\gamma_{1}, \cdots, J\gamma_{p}, {\cal C}_{+}{\cal P} {\cal K}, J {\cal C}_{+}{\cal P} {\cal K} \}$ & $Cl_{p+1, 0} \rightarrow Cl_{p+2, 0}$ & ${\cal R}_{p+3}$ \\ \hline
	\multirow{3}{*}{${\cal P}\text{CII}^{\dag}$} 
	& P & ~$\{ \gamma_{1}, \cdots, \gamma_{p}, {\sf \Sigma}, {\sf \Sigma} ({\sf C}_{+}{\sf P} {\cal K}), J{\sf \Sigma} ({\sf C}_{+}{\sf P} {\cal K}), ({\sf P}{\sf T}_{-}{\cal K}) ({\sf C}_{+}{\sf P} {\cal K}) \}$~ & $Cl_{0, p+3} \rightarrow Cl_{0, p+4}$ & ${\cal R}_{p+3}$ \\ 
	& \Lr & ~$\{ {\cal P}{\cal T}_{-}{\cal K}, J{\cal P}{\cal T}_{-}{\cal K}, J ({\cal P}{\cal T}_{-}{\cal K}) ({\cal C}_{+}{\cal P} {\cal K}); \gamma_{1}, \cdots, \gamma_{p} \}$~ & $Cl_{3, p-1} \rightarrow Cl_{3, p}$ & ${\cal R}_{p+4}$ \\
	& \Li & $\{ J\gamma_{1}, \cdots, J\gamma_{p}, {\cal C}_{+}{\cal P} {\cal K}, J {\cal C}_{+}{\cal P} {\cal K} \} \otimes \{ ({\cal P}{\cal T}_{-}{\cal K}) ({\cal C}_{+}{\cal P} {\cal K}) \}$ & $Cl_{p+1, 0} \times Cl_{p+1, 0} \rightarrow Cl_{p+2, 0} \times Cl_{p+2, 0}$ & ${\cal R}_{p+3} \times {\cal R}_{p+3}$ \\ \hline
	\multirow{3}{*}{${\cal P}\text{C}^{\dag}$} 
	& P & $\{ J{\sf \Sigma}, {\sf P}{\sf T}_{-}{\cal K}, J{\sf P}{\sf T}_{-}{\cal K}; \gamma_{1}, \cdots, \gamma_{p} \}$ & $Cl_{3, p-1} \rightarrow Cl_{3, p}$ & ${\cal R}_{p+4}$ \\ 
	& \Lr & $\{ {\cal P}{\cal T}_{-}{\cal K}, J{\cal P}{\cal T}_{-}{\cal K}; \gamma_{1}, \cdots, \gamma_{p} \}$ & $Cl_{2, p-1} \rightarrow Cl_{2, p}$ & ${\cal R}_{p+5}$ \\
	& \Li & $\{ J \gamma_{1}, \cdots, J\gamma_{p}, {\cal P}{\cal T}_{-}{\cal K}, J{\cal P}{\cal T}_{-}{\cal K} \}$ & $Cl_{p+1, 0} \rightarrow Cl_{p+2, 0}$ & ${\cal R}_{p+4}$ \\ \hline
	\multirow{3}{*}{${\cal P}\text{CI}^{\dag}$} 
	& P & ~$\{ {\sf \Sigma} ({\sf C}_{+}{\sf P} {\cal K}), J{\sf \Sigma} ({\sf C}_{+}{\sf P} {\cal K}), ({\sf P}{\sf T}_{-}{\cal K}) ({\sf C}_{+}{\sf P} {\cal K}); \gamma_{1}, \cdots, \gamma_{p}, {\sf \Sigma} \}$~ & $Cl_{3, p} \rightarrow Cl_{3, p+1}$ & ${\cal R}_{p+5}$ \\ 
	& \Lr & ~$\{ {\cal P}{\cal T}_{-}{\cal K}, J{\cal P}{\cal T}_{-}{\cal K}; \gamma_{1}, \cdots, \gamma_{p}, J ({\cal P}{\cal T}_{-}{\cal K}) ({\cal C}_{+}{\cal P} {\cal K}) \}$~ & $Cl_{2, p} \rightarrow Cl_{2, p+1}$ & ${\cal R}_{p+6}$ \\
	& \Li & $\{ J\gamma_{1}, \cdots, J\gamma_{p}; {\cal C}_{+}{\cal P} {\cal K}, J {\cal C}_{+}{\cal P} {\cal K} \} \otimes \{ ({\sf P}{\sf T}_{-}{\cal K}) ({\cal C}_{+}{\cal P} {\cal K}) \}$ & $Cl_{p+1} \rightarrow Cl_{p+2}$ & ${\cal C}_{p+1}$ \\ \hline \hline % real AZ^{\dag}
	\multirow{3}{*}{~AIII + ${\cal S}_{+}$~} 
	& ~P~ & $\{ \gamma_{1}, \cdots, \gamma_{p}, {\sf \Sigma}, {\sf \Gamma}, {\sf \Sigma \Gamma S} \}$ & $Cl_{p+2} \rightarrow Cl_{p+3}$ & ${\cal C}_{p}$ \\ 
	& ~\Lr~ & $\{ \gamma_{1}, \cdots, \gamma_{p}, \Gamma \} \otimes \{ \Gamma {\cal S} \}$ & $Cl_{p} \times Cl_{p} \rightarrow Cl_{p+1}  \times Cl_{p+1}$ & ${\cal C}_{p} \times {\cal C}_{p}$ \\
	& ~\Li~ & $\{ \gamma_{1}, \cdots, \gamma_{p}, {\cal S} \} \otimes \{ \Gamma \}$ & $Cl_{p} \times Cl_{p} \rightarrow Cl_{p+1}  \times Cl_{p+1}$ & ${\cal C}_{p} \times {\cal C}_{p}$ \\ \hline 
	\multirow{2}{*}{A + ${\cal S}$} 
	& P & $\{ \gamma_{1}, \cdots, \gamma_{p}, {\sf \Sigma} \} \otimes \{ {\sf S} \}$ & ~~$Cl_{p} \times Cl_{p} \rightarrow Cl_{p+1} \times Cl_{p+1}$~~ & ${\cal C}_{p} \times {\cal C}_{p}$ \\ 
	& L & $\{ \gamma_{1}, \cdots, \gamma_{p}, {\cal S} \}$ & $Cl_{p} \rightarrow Cl_{p+1}$ & ${\cal C}_{p}$ \\ \hline
	\multirow{3}{*}{~AIII + ${\cal S}_{-}$~} 
	& P & $\{ \gamma_{1}, \cdots, \gamma_{p}, {\sf \Sigma}, {\sf \Gamma} \} \otimes \{ {\sf \Sigma S} \}$ & ~~$Cl_{p+1} \times Cl_{p+1} \rightarrow Cl_{p+2} \times Cl_{p+2}$~~ & ~${\cal C}_{p+1} \times {\cal C}_{p+1}$~ \\ 
	& \Lr & $\{ \gamma_{1}, \cdots, \gamma_{p}, \Gamma, {\cal S} \}$ & $Cl_{p+1} \rightarrow Cl_{p+2}$ & ${\cal C}_{p+1}$ \\
	& \Li & $\{ \gamma_{1}, \cdots, \gamma_{p}, {\cal S}, \Gamma{\cal S} \}$ & $Cl_{p+1} \rightarrow Cl_{p+2}$ & ${\cal C}_{p+1}$ \\ \hline \hline % complex AZ + SLS
	\multirow{3}{*}{${\cal P}$BDI + ${\cal S}_{++}$} 
	& P & $\{ J ({\sf CP}{\cal K}) ({\sf PT}{\cal K}), ({\sf CP}{\cal K}) ({\sf PT}{\cal K}) {\sf \Sigma S}; \gamma_{1}, \cdots, \gamma_{p}, {\sf \Sigma}, {\sf CP}{\cal K}, J{\sf CP}{\cal K}\}$ & $Cl_{2, p+2} \rightarrow Cl_{2, p+3}$ & ${\cal R}_{p}$ \\ 
	& \Lr & ~$\{ J ({\cal CPK}) ({\cal PTK}); \gamma_{1}, \cdots, \gamma_{p}, {\cal CPK}, J{\cal CPK} \} \otimes \{ ({\cal CPK}) ({\cal PTK}) {\cal S}\}$~ & ~$Cl_{1, p+1} \times Cl_{1, p+1} \rightarrow Cl_{1, p+2} \times Cl_{1, p+2}$~ & ${\cal R}_{p} \times {\cal R}_{p}$ \\
	& \Li & $\{ J{\cal S}; \gamma_{1}, \cdots, \gamma_{p}, {\cal CPK}, J{\cal CPK}\} \otimes \{ ({\cal CPK}) ({\cal PTK}) \}$ & $Cl_{1, p+1} \times Cl_{1, p+1} \rightarrow Cl_{1, p+2} \times Cl_{1, p+2}$ & ${\cal R}_{p} \times {\cal R}_{p}$ \\ \hline
	\multirow{3}{*}{${\cal P}$DIII + ${\cal S}_{--}$} 
	& P & $\{ J ({\sf CP}{\cal K}) ({\sf PT}{\cal K}) {\sf \Sigma S}; \gamma_{1}, \cdots, \gamma_{p}, {\sf \Sigma}, {\sf CP}{\cal K}, J{\sf CP}{\cal K}, J ({\sf CP}{\cal K}) ({\sf PT}{\cal K}) \}$ & $Cl_{1, p+3} \rightarrow Cl_{1, p+4}$ & ${\cal R}_{p+2}$ \\ 
	& \Lr & $\{ \gamma_{1}, \cdots, \gamma_{p}, {\cal CPK}, J{\cal CPK}, J ({\cal CPK}) ({\cal PTK}) \} \otimes \{ ({\cal CPK}) ({\cal PTK}) {\cal S} \}$ & $Cl_{0, p+2} \times Cl_{0, p+2} \rightarrow Cl_{0, p+3} \times Cl_{0, p+3}$ & ${\cal R}_{p+2} \times {\cal R}_{p+2}$ \\
	& \Li & $\{ \gamma_{1}, \cdots, \gamma_{p}, {\cal CPK}, J{\cal CPK}, {\cal S} \} \otimes \{ ({\cal CPK}) ({\cal PTK}) \}$ & $Cl_{p+2} \rightarrow Cl_{p+3}$ & ${\cal C}_{p}$ \\ \hline
	\multirow{3}{*}{${\cal P}$CII + ${\cal S}_{++}$} 
	& P & $\{ {\sf CP}{\cal K}, J{\sf CP}{\cal K}, J ({\sf CP}{\cal K}) ({\sf PT}{\cal K}), ({\sf CP}{\cal K}) ({\sf PT}{\cal K}) {\sf \Sigma S}; \gamma_{1}, \cdots, \gamma_{p}, {\sf \Sigma}\}$ & $Cl_{4, p} \rightarrow Cl_{4, p+1}$ & ${\cal R}_{p+4}$ \\ 
	& \Lr & ~$\{ {\cal CPK}, J{\cal CPK}, J ({\cal CPK}) ({\cal PTK}); \gamma_{1}, \cdots, \gamma_{p} \} \otimes \{ ({\cal CPK}) ({\cal PTK}) {\cal S}\}$~ & ~$Cl_{3, p-1} \times Cl_{3, p-1} \rightarrow Cl_{3, p} \times Cl_{3, p}$~ & ${\cal R}_{p+4} \times {\cal R}_{p+4}$ \\
	& \Li & $\{ {\cal CPK}, J{\cal CPK}, J{\cal S}; \gamma_{1}, \cdots, \gamma_{p}\} \otimes \{ ({\cal CPK}) ({\cal PTK}) \}$ & $Cl_{3, p-1} \times Cl_{3, p-1} \rightarrow Cl_{3, p} \times Cl_{3, p}$ & ${\cal R}_{p+4} \times {\cal R}_{p+4}$ \\ \hline
	\multirow{3}{*}{${\cal P}$CI + ${\cal S}_{--}$} 
	& P & $\{ {\sf CP}{\cal K}, J{\sf CP}{\cal K}, J ({\sf CP}{\cal K}) ({\sf PT}{\cal K}) {\sf \Sigma S}; \gamma_{1}, \cdots, \gamma_{p}, {\sf \Sigma}, J ({\sf CP}{\cal K}) ({\sf PT}{\cal K}) \}$ & $Cl_{3, p+1} \rightarrow Cl_{3, p+1}$ & ${\cal R}_{p+6}$ \\ 
	& \Lr & $\{ {\cal CPK}, J{\cal CPK}; \gamma_{1}, \cdots, \gamma_{p}, J ({\cal CPK}) ({\cal PTK}) \} \otimes \{ ({\cal CPK}) ({\cal PTK}) {\cal S} \}$ & $Cl_{2, p} \times Cl_{2, p} \rightarrow Cl_{2, p+1} \times Cl_{2, p+1}$ & ~~${\cal R}_{p+6} \times {\cal R}_{p+6}$~~ \\
	& \Li & $\{ {\cal CPK}, J{\cal CPK}; \gamma_{1}, \cdots, \gamma_{p}, {\cal S} \} \otimes \{ ({\cal CPK}) ({\cal PTK}) \}$ & $Cl_{p+2} \rightarrow Cl_{p+3}$ & ${\cal C}_{p}$ \\ \hline \hline % real AZ + SLS #1
	\multirow{3}{*}{${\cal P}$AI + ${\cal S}_{-}$} 
	& P & ~~$\{ J\gamma_{1}, \cdots, J\gamma_{p}, J{\sf \Sigma}; {\sf PT}{\cal K}, J{\sf PT}{\cal K} \} \otimes \{ J{\sf \Sigma S} \}$~~ & $Cl_{p+2} \rightarrow Cl_{p+3}$ & ${\cal C}_{p}$ \\ 
	& \Lr & $\{ J\gamma_{1}, \cdots, J\gamma_{p}; {\cal PTK}, J{\cal PTK}, {\cal S} \}$ & $Cl_{p-1, 3} \rightarrow Cl_{p, 3}$ & ${\cal R}_{p+6}$ \\
	& \Li & $\{ \gamma_{1}, \cdots, \gamma_{p}, {\cal PTK}, J{\cal PTK}, {\cal S} \}$ & $Cl_{0, p+2} \rightarrow Cl_{0, p+3}$ & ${\cal R}_{p+2}$ \\ \hline
	\multirow{3}{*}{${\cal P}$BDI + ${\cal S}_{-+}$} 
	& P & $\{ J ({\sf CP}{\cal K}) ({\sf PT}{\cal K}); \gamma_{1}, \cdots, \gamma_{p}, {\sf \Sigma}, {\sf CP}{\cal K}, J{\sf CP}{\cal K}\} \otimes \{ J{\sf \Sigma S} \}$ & $Cl_{p+3} \rightarrow Cl_{p+4}$ & ${\cal C}_{p+1}$ \\ 
	& \Lr & $\{ J{\cal S}, J ({\cal CPK}) ({\cal PTK}); \gamma_{1}, \cdots, \gamma_{p}, {\cal CPK}, J{\cal CPK} \}$ & $Cl_{2, p+1} \rightarrow Cl_{2, p+2}$ & ${\cal R}_{p+7}$ \\
	& \Li & $\{ J{\cal S}; \gamma_{1}, \cdots, \gamma_{p}, {\cal CPK}, J {\cal CPK}, J ({\cal CPK}) ({\cal PTK}) {\cal S} \}$ & $Cl_{1, p+2} \rightarrow Cl_{1, p+3}$ & ${\cal R}_{p+1}$ \\ \hline
	\multirow{2}{*}{${\cal P}$D + ${\cal S}_{+}$} 
	& P & ~~$\{ \gamma_{1}, \cdots, \gamma_{p}, {\sf \Sigma}, {\sf CP}{\cal K}, J{\sf CP}{\cal K} \} \otimes \{ J{\sf \Sigma S} \}$~~ & $Cl_{p+2} \rightarrow Cl_{p+3}$ & ${\cal C}_{p}$ \\ 
	& L & $\{ J{\cal S}; \gamma_{1}, \cdots, \gamma_{p}, {\cal CPK}, J{\cal CPK} \}$ & $Cl_{1, p+1} \rightarrow Cl_{1, p+2}$ & ${\cal R}_{p}$ \\ \hline
	\multirow{3}{*}{${\cal P}$DIII + ${\cal S}_{-+}$} 
	& P & $\{ \gamma_{1}, \cdots, \gamma_{p}, {\sf \Sigma}, {\sf CP}{\cal K}, J{\sf CP}{\cal K}, J ({\sf CP}{\cal K}) ({\sf PT}{\cal K}) \} \otimes \{ J{\sf \Sigma S} \}$ & $Cl_{p+3} \rightarrow Cl_{p+4}$ & ${\cal C}_{p+1}$ \\ 
	& \Lr & $\{ J{\cal S}; \gamma_{1}, \cdots, \gamma_{p}, {\cal CPK}, J{\cal CPK}, J ({\cal CPK}) ({\cal PTK}) \}$ & $Cl_{1, p+2} \rightarrow Cl_{1, p+3}$ & ${\cal R}_{p+1}$ \\
	& \Li & $\{ J{\cal S}, J ({\cal CPK}) ({\cal PTK}) {\cal S}; \gamma_{1}, \cdots, \gamma_{p}, {\cal CPK}, J {\cal CPK} \}$ & $Cl_{2, p+1} \rightarrow Cl_{2, p+2}$ & ${\cal R}_{p+7}$ \\ \hline
	\multirow{3}{*}{${\cal P}$AII + ${\cal S}_{-}$} 
	& P & ~~$\{ J\gamma_{1}, \cdots, J\gamma_{p}, J{\sf \Sigma}, {\sf PT}{\cal K}, J{\sf PT}{\cal K} \} \otimes \{ J{\sf \Sigma S} \}$~~ & $Cl_{p+2} \rightarrow Cl_{p+3}$ & ${\cal C}_{p}$ \\ 
	& \Lr & $\{ J\gamma_{1}, \cdots, J\gamma_{p}, {\cal PTK}, J{\cal PTK}; {\cal S} \}$ & $Cl_{p+1, 1} \rightarrow Cl_{p+2, 1}$ & ${\cal R}_{p+2}$ \\
	& \Li & $\{ {\cal PTK}, J{\cal PTK}; \gamma_{1}, \cdots, \gamma_{p}, {\cal S} \}$ & $Cl_{2, p} \rightarrow Cl_{2, p+1}$ & ${\cal R}_{p+6}$ \\ \hline
	\multirow{3}{*}{${\cal P}$CII + ${\cal S}_{-+}$} 
	& P & $\{ {\sf CP}{\cal K}, J{\sf CP}{\cal K}, J ({\sf CP}{\cal K}) ({\sf PT}{\cal K}); \gamma_{1}, \cdots, \gamma_{p}, {\sf \Sigma} \} \otimes \{ J{\sf \Sigma S} \}$ & $Cl_{p+3} \rightarrow Cl_{p+4}$ & ${\cal C}_{p+1}$ \\ 
	& \Lr & $\{ {\cal CPK}, J{\cal CPK}, J ({\cal CPK}) ({\cal PTK}), J{\cal S}; \gamma_{1}, \cdots, \gamma_{p} \}$ & $Cl_{4, p-1} \rightarrow Cl_{4, p}$ & ${\cal R}_{p+3}$ \\
	& \Li & $\{ {\cal CPK}, J {\cal CPK}, J{\cal S}; \gamma_{1}, \cdots, \gamma_{p}, J ({\cal CPK}) ({\cal PTK}) {\cal S} \}$ & $Cl_{3, p} \rightarrow Cl_{3, p+1}$ & ${\cal R}_{p+5}$ \\ \hline
	\multirow{2}{*}{${\cal P}$C + ${\cal S}_{+}$} 
	& P & ~~$\{ {\sf CP}{\cal K}, J{\sf CP}{\cal K}; \gamma_{1}, \cdots, \gamma_{p}, {\sf \Sigma} \} \otimes \{ J{\sf \Sigma S} \}$~~ & $Cl_{p+2} \rightarrow Cl_{p+3}$ & ${\cal C}_{p}$ \\ 
	& L & $\{ {\cal CPK}, J{\cal CPK}, J{\cal S}; \gamma_{1}, \cdots, \gamma_{p} \}$ & $Cl_{3, p-1} \rightarrow Cl_{3, p}$ & ${\cal R}_{p+4}$ \\ \hline 
	\multirow{3}{*}{${\cal P}$CI + ${\cal S}_{-+}$} 
	& P & $\{ {\sf CP}{\cal K}, J{\sf CP}{\cal K}; \gamma_{1}, \cdots, \gamma_{p}, {\sf \Sigma}, J ({\sf CP}{\cal K}) ({\sf PT}{\cal K}) \} \otimes \{ J{\sf \Sigma S} \}$ & $Cl_{p+3} \rightarrow Cl_{p+4}$ & ${\cal C}_{p+1}$ \\ 
	& \Lr & $\{ {\cal CPK}, J{\cal CPK}, J{\cal S}; \gamma_{1}, \cdots, \gamma_{p}, J ({\cal CPK}) ({\cal PTK}) \}$ & $Cl_{3, p} \rightarrow Cl_{3, p+1}$ & ${\cal R}_{p+5}$ \\
	& \Li & $\{ {\cal CPK}, J {\cal CPK}, J{\cal S}, J ({\cal CPK}) ({\cal PTK}) {\cal S}; \gamma_{1}, \cdots, \gamma_{p} \}$ & $Cl_{4, p-1} \rightarrow Cl_{4, p}$ & ${\cal R}_{p+3}$ \\ \hline \hline % real AZ + SLS #2
   \multicolumn{5}{c}{continued on next page}
  \end{tabular}
\end{table}

\begin{table}[H]
	\centering
	\footnotesize
     \begin{tabular}{ccccc} % \hline \hline
%    ~Symmetry class~ & ~Gap~ & ~Generator~ & ~Extension~ & Classifying space \\ \hline
    \multicolumn{5}{c}{TABLE \ref{tab: extension} --- continued} \\ \hline \hline
%    ~Symmetry class~ & ~Gap~ & ~Generator~ & ~Extension~ & Classifying space \\ \hline
  	\multirow{3}{*}{${\cal P}$BDI + ${\cal S}_{--}$} 
	& P & $\{ J ({\sf CP}{\cal K}) ({\sf PT}{\cal K}); \gamma_{1}, \cdots, \gamma_{p}, {\sf \Sigma}, {\sf CP}{\cal K}, J{\sf CP}{\cal K}, J ({\sf CP}{\cal K}) ({\sf PT}{\cal K}) {\sf \Sigma S}\}$ & $Cl_{1, p+3} \rightarrow Cl_{1, p+4}$ & ${\cal R}_{p+2}$ \\ 
	& \Lr & $\{ J ({\cal CPK}) ({\cal PTK}); \gamma_{1}, \cdots, \gamma_{p}, {\cal CPK}, J{\cal CPK} \} \otimes \{ ({\cal CPK}) ({\cal PTK}) {\cal S} \}$ & $Cl_{p+2} \rightarrow Cl_{p+3}$ & ${\cal C}_{p}$ \\
	& \Li & $\{ \gamma_{1}, \cdots, \gamma_{p}, {\cal CPK}, J{\cal CPK}, {\cal S} \} \otimes \{ ({\cal CPK}) ({\cal PTK}) \}$ & $Cl_{0, p+2} \times Cl_{0, p+2} \rightarrow Cl_{0, p+3} \times Cl_{0, p+3}$ & ${\cal R}_{p+2} \times {\cal R}_{p+2}$ \\ \hline
	\multirow{3}{*}{${\cal P}$DIII + ${\cal S}_{++}$} 
	& P & $\{ \gamma_{1}, \cdots, \gamma_{p}, {\sf \Sigma}, {\sf CP}{\cal K}, J{\sf CP}{\cal K}, J ({\sf CP}{\cal K}) ({\sf PT}{\cal K}), ({\sf CP}{\cal K}) ({\sf PT}{\cal K}) {\sf \Sigma S} \}$ & $Cl_{0, p+4} \rightarrow Cl_{0, p+5}$ & ${\cal R}_{p+4}$ \\ 
	& \Lr & ~$\{ \gamma_{1}, \cdots, \gamma_{p}, {\cal CPK}, J{\cal CPK}, J ({\cal CPK}) ({\cal PTK}) \} \otimes \{ ({\cal CPK}) ({\cal PTK}) {\cal S}\}$~ & $Cl_{p+2} \rightarrow Cl_{p+3}$ & ${\cal C}_{p}$ \\
	& \Li & $\{ J{\cal S}; \gamma_{1}, \cdots, \gamma_{p}, {\cal CPK}, J{\cal CPK} \} \otimes \{ ({\cal CPK}) ({\cal PTK}) \}$ & $Cl_{p+2} \rightarrow Cl_{p+3}$ & ${\cal C}_{p}$ \\ \hline
	\multirow{3}{*}{${\cal P}$CII + ${\cal S}_{--}$} 
	& P & $\{ {\sf CP}{\cal K}, J{\sf CP}{\cal K}, J ({\sf CP}{\cal K}) ({\sf PT}{\cal K}); \gamma_{1}, \cdots, \gamma_{p}, {\sf \Sigma}, J ({\sf CP}{\cal K}) ({\sf PT}{\cal K}) {\sf \Sigma S} \}$ & $Cl_{3, p+1} \rightarrow Cl_{3, p+1}$ & ${\cal R}_{p+6}$ \\ 
	& \Lr & $\{ {\cal CPK}, J{\cal CPK}, J ({\cal CPK}) ({\cal PTK}); \gamma_{1}, \cdots, \gamma_{p} \} \otimes \{ ({\cal CPK}) ({\cal PTK}) {\cal S} \}$ & $Cl_{p+2} \rightarrow Cl_{p+3}$ & ${\cal C}_{p}$ \\
	& \Li & $\{ {\cal CPK}, J{\cal CPK}; \gamma_{1}, \cdots, \gamma_{p}, {\cal S} \} \otimes \{ ({\cal CPK}) ({\cal PTK}) \}$ & $Cl_{2, p} \times Cl_{2, p} \rightarrow Cl_{2, p+1} \times Cl_{2, p+1}$ & ${\cal R}_{p+6} \times {\cal R}_{p+6}$ \\ \hline
	\multirow{3}{*}{${\cal P}$CI + ${\cal S}_{++}$} 
	& P & $\{ {\sf CP}{\cal K}, J{\sf CP}{\cal K}; \gamma_{1}, \cdots, \gamma_{p}, {\sf \Sigma}, J ({\sf CP}{\cal K}) ({\sf PT}{\cal K}), ({\sf CP}{\cal K}) ({\sf PT}{\cal K}) {\sf \Sigma S} \}$ & $Cl_{2, p+2} \rightarrow Cl_{2, p+3}$ & ${\cal R}_{p}$ \\ 
	& \Lr & ~$\{ {\cal CPK}, J{\cal CPK}; \gamma_{1}, \cdots, \gamma_{p}, J ({\cal CPK}) ({\cal PTK}) \} \otimes \{ ({\cal CPK}) ({\cal PTK}) {\cal S}\}$~ & $Cl_{p+2} \rightarrow Cl_{p+3}$ & ${\cal C}_{p}$ \\
	& \Li & $\{ {\cal CPK}, J{\cal CPK}, J{\cal S}; \gamma_{1}, \cdots, \gamma_{p} \} \otimes \{ ({\cal CPK}) ({\cal PTK}) \}$ & $Cl_{p+2} \rightarrow Cl_{p+3}$ & ${\cal C}_{p}$ \\ \hline \hline % real AZ + SLS #3
	\multirow{3}{*}{${\cal P}$AI + ${\cal S}_{+}$} 
	& ~P~ & ~~$\{ J\gamma_{1}, \cdots, J\gamma_{p}, J{\sf \Sigma}; {\sf PT}{\cal K}, J{\sf PT}{\cal K} \} \otimes \{ {\sf \Sigma S} \}$~~ & ~~$Cl_{p, 2} \times Cl_{p, 2} \rightarrow Cl_{p+1, 2} \times Cl_{p+1, 2}$~~ & ${\cal R}_{p} \times {\cal R}_{p}$ \\ 
	& ~\Lr~ & $\{ J\gamma_{1}, \cdots, J\gamma_{p}, {\cal S}; {\cal PTK}, J{\cal PTK} \}$ & $Cl_{p, 2} \rightarrow Cl_{p+1, 2}$ & ${\cal R}_{p}$ \\
	& ~\Li~ & $\{ J{\cal S}; \gamma_{1}, \cdots, \gamma_{p}, {\cal PTK}, J{\cal PTK} \}$ & $Cl_{1, p+1} \rightarrow Cl_{1, p+2}$ & ${\cal R}_{p}$ \\ \hline
	\multirow{3}{*}{${\cal P}$BDI + ${\cal S}_{+-}$} 
	& P & $\{ J ({\sf CP}{\cal K}) ({\sf PT}{\cal K}); \gamma_{1}, \cdots, \gamma_{p}, {\sf \Sigma}, {\sf CP}{\cal K}, J{\sf CP}{\cal K}\} \otimes \{ {\sf \Sigma S} \}$ & $Cl_{1, p+2} \times Cl_{1, p+2} \rightarrow Cl_{1, p+3} \times Cl_{1, p+3}$ & ~~${\cal R}_{p+1} \times {\cal R}_{p+1}$~~ \\ 
	& \Lr & $\{ J ({\cal CPK}) ({\cal PTK}); \gamma_{1}, \cdots, \gamma_{p}, {\cal CPK}, J{\cal CPK}, {\cal S} \}$ & $Cl_{1, p+2} \rightarrow Cl_{1, p+3}$ & ${\cal R}_{p+1}$ \\
	& \Li & $\{ ({\cal CPK}) ({\cal PTK}) {\cal S}; \gamma_{1}, \cdots, \gamma_{p}, {\cal CPK}, J{\cal CPK}, {\cal S}\}$ & $Cl_{1, p+2} \rightarrow Cl_{1, p+3}$ & ${\cal R}_{p+1}$ \\ \hline
	\multirow{2}{*}{${\cal P}$D + ${\cal S}_{-}$} 
	& P & ~~$\{ \gamma_{1}, \cdots, \gamma_{p}, {\sf \Sigma}, {\sf CP}{\cal K}, J{\sf CP}{\cal K} \} \otimes \{ {\sf \Sigma S} \}$~~ & ~~$Cl_{0, p+2} \times Cl_{0, p+2} \rightarrow Cl_{0, p+3} \times Cl_{0, p+3}$~~ & ${\cal R}_{p+2} \times {\cal R}_{p+2}$ \\ 
	& L & $\{ \gamma_{1}, \cdots, \gamma_{p}, {\cal CPK}, J{\cal CPK}, {\cal S} \}$ & $Cl_{0, p+2} \rightarrow Cl_{0, p+3}$ & ${\cal R}_{p+2}$ \\ \hline
	\multirow{3}{*}{${\cal P}$DIII + ${\cal S}_{+-}$} 
	& P & $\{ \gamma_{1}, \cdots, \gamma_{p}, {\sf \Sigma}, {\sf CP}{\cal K}, J{\sf CP}{\cal K}, J ({\sf CP}{\cal K}) ({\sf PT}{\cal K}) \} \otimes \{ {\sf \Sigma S} \}$ & $Cl_{0, p+3} \times Cl_{0, p+3} \rightarrow Cl_{0, p+4} \times Cl_{0, p+4}$ & ${\cal R}_{p+3} \times {\cal R}_{p+3}$ \\ 
	& \Lr & $\{ \gamma_{1}, \cdots, \gamma_{p}, {\cal CPK}, J{\cal CPK}, J ({\cal CPK}) ({\cal PTK}), {\cal S} \}$ & $Cl_{0, p+3} \rightarrow Cl_{0, p+4}$ & ${\cal R}_{p+3}$ \\
	& \Li & $\{ \gamma_{1}, \cdots, \gamma_{p}, {\cal CPK}, J{\cal CPK}, {\cal S}, ({\cal CPK}) ({\cal PTK}) {\cal S} \}$ & $Cl_{0, p+3} \rightarrow Cl_{0, p+4}$ & ${\cal R}_{p+3}$ \\ \hline
	\multirow{3}{*}{${\cal P}$AII + ${\cal S}_{+}$} 
	& P & ~~$\{ J\gamma_{1}, \cdots, J\gamma_{p}, J{\sf \Sigma}, {\sf PT}{\cal K}, J{\sf PT}{\cal K} \} \otimes \{ {\sf \Sigma S} \}$~~ & ~~$Cl_{p+2, 0} \times Cl_{p+2, 0} \rightarrow Cl_{p+3, 0} \times Cl_{p+3, 0}$~~ & ${\cal R}_{p+4} \times {\cal R}_{p+4}$ \\ 
	& \Lr & $\{ J\gamma_{1}, \cdots, J\gamma_{p}, {\cal PTK}, J{\cal PTK}, {\cal S} \}$ & $Cl_{p+2, 0} \rightarrow Cl_{p+3, 0}$ & ${\cal R}_{p+4}$ \\
	& \Li & $\{ {\cal PTK}, J{\cal PTK}, J{\cal S}; \gamma_{1}, \cdots, \gamma_{p} \}$ & $Cl_{3, p-1} \rightarrow Cl_{3, p}$ & ${\cal R}_{p+4}$ \\ \hline
	\multirow{3}{*}{${\cal P}$CII + ${\cal S}_{+-}$} 
	& P & $\{ {\sf CP}{\cal K}, J{\sf CP}{\cal K}, J ({\sf CP}{\cal K}) ({\sf PT}{\cal K}); \gamma_{1}, \cdots, \gamma_{p}, {\sf \Sigma} \} \otimes \{ {\sf \Sigma S} \}$ & $Cl_{3, p} \times Cl_{3, p} \rightarrow Cl_{3, p+1} \times Cl_{3, p+1}$ & ${\cal R}_{p+5} \times {\cal R}_{p+5}$ \\ 
	& \Lr & $\{ {\cal CPK}, J{\cal CPK}, J ({\cal CPK}) ({\cal PTK}); \gamma_{1}, \cdots, \gamma_{p}, {\cal S} \}$ & $Cl_{3, p} \rightarrow Cl_{3, p+1}$ & ${\cal R}_{p+5}$ \\
	& \Li & $\{ {\cal CPK}, J{\cal CPK}, ({\cal CPK}) ({\cal PTK}) {\cal S}; \gamma_{1}, \cdots, \gamma_{p}, {\cal S} \}$ & $Cl_{3, p} \rightarrow Cl_{3, p+1}$ & ${\cal R}_{p+5}$ \\ \hline
	\multirow{2}{*}{${\cal P}$C + ${\cal S}_{-}$} 
	& P & ~~$\{ {\sf CP}{\cal K}, J{\sf CP}{\cal K}; \gamma_{1}, \cdots, \gamma_{p}, {\sf \Sigma} \} \otimes \{ {\sf \Sigma S} \}$~~ & ~~$Cl_{2, p} \times Cl_{2, p} \rightarrow Cl_{2, p+1} \times Cl_{2, p+1}$~~ & ${\cal R}_{p+6} \times {\cal R}_{p+6}$ \\ 
	& L & $\{ {\cal CPK}, J{\cal CPK}; \gamma_{1}, \cdots, \gamma_{p}, {\cal S} \}$ & $Cl_{2, p} \rightarrow Cl_{2, p+1}$ & ${\cal R}_{p+6}$ \\ \hline
	\multirow{3}{*}{${\cal P}$CI + ${\cal S}_{+-}$} 
	& P & $\{ {\sf CP}{\cal K}, J{\sf CP}{\cal K}; \gamma_{1}, \cdots, \gamma_{p}, {\sf \Sigma}, J ({\sf CP}{\cal K}) ({\sf PT}{\cal K}) \} \otimes \{ {\sf \Sigma S} \}$ & $Cl_{2, p+1} \times Cl_{2, p+1} \rightarrow Cl_{2, p+2} \times Cl_{2, p+2}$ & ${\cal R}_{p+7} \times {\cal R}_{p+7}$ \\ 
	& \Lr & $\{ {\cal CPK}, J{\cal CPK}; \gamma_{1}, \cdots, \gamma_{p}, J ({\cal CPK}) ({\cal PTK}), {\cal S} \}$ & $Cl_{2, p+1} \rightarrow Cl_{2, p+2}$ & ${\cal R}_{p+7}$ \\
	& \Li & $\{ {\cal CPK}, J{\cal CPK}; \gamma_{1}, \cdots, \gamma_{p}, {\cal S}, ({\cal CPK}) ({\cal PTK}) {\cal S} \}$ & $Cl_{2, p+1} \rightarrow Cl_{2, p+2}$ & ${\cal R}_{p+7}$ \\ \hline \hline % real AZ + SLS #4
  \end{tabular}
\end{table}

%%%%% classification table %%%%%
\begin{table}[H]
	\centering
	\caption{Classification of topologically stable exceptional points in the previous works~\cite{Berry-04, Esaki-11, Zhen-15, Gonzalez-16, Leykam-17, Xu-17, Gonzalez-17, Chernodub-17, Zyuzin-18, Cerjan-18, Zhou-18-exp, Molina-18, Shen-18, Shen-Fu-18, Carlstrom-18, Moors-19, Okugawa-19, Budich-19, Yang-19, Zhou-19, Wang-19, Yoshida-19, Carlstrom-19, Zyuzin-19, Papaj-19, Kozii-17, Cerjan-18-exp, Luo-18, Lee-18-tidal, Chernodub-19, Bergholtz-19}. The codimension $p$ is defined as $p := d-d_{\rm EP}$ with the spatial dimension $d$ and the dimension $d_{\rm EP}$ of the gapless region; exceptional points, lines, and surfaces are described by $d_{\rm EP} = 0, 1, 2$, respectively. Complex-energy gaps have two distinct types, a point (P) or line (L) gap, and the subscript of L specifies a line gap for the real or imaginary part of the complex spectrum. The sign of $\PT$ symmetry means $(\PT)\,(\PT)^{*} = +1$.\\}
	\label{tab: previous works}
     \begin{tabular}{ccccc} \hline \hline
    ~~Symmetry~~ & ~Gap~ & ~~~$p=1$~~~ & ~~~$p=2$~~~ & ~~~$p=3$~~~ \\ \hline
    \multirow{2}{*}{No} 
    & P & $0$ & \Z~\cite{Berry-04, Xu-17, Gonzalez-16, Gonzalez-17, Zyuzin-18, Cerjan-18, Zhou-18-exp, Molina-18, Shen-18, Shen-Fu-18, Carlstrom-18, Yang-19, Carlstrom-19, Zyuzin-19, Papaj-19, Kozii-17, Cerjan-18-exp, Luo-18, Lee-18-tidal, Chernodub-19, Bergholtz-19} & $0$ \\ 
    & L & \Z & $0$ & \Z~\cite{Xu-17, Cerjan-18, Cerjan-18-exp, Chernodub-19} \\ \hline
    \multirow{3}{*}{Chiral} 
    & P & \Z~\cite{Esaki-11, Budich-19, Yoshida-19} & $0$ & \Z~[Eq.~(\ref{eq: CS-p3})] \\ 
    & \Lr & $0$ & \Z & $0$ \\
    & \Li & \Z\,$\oplus$\,\Z & $0$ & \Z\,$\oplus$\,\Z \\ \hline
    \multirow{2}{*}{Sublattice} 
    & P & $0$ & \Z\,$\oplus$\,\Z~\cite{Leykam-17, Wang-19, Moors-19} & $0$ \\ 
    & L & $0$ & \Z & $0$ \\ \hline
    \multirow{3}{*}{$\PT$, $+1$} 
    & P & \Zt~\cite{Zhen-15, Chernodub-17, Okugawa-19, Zhou-19} & \Zt & $0$ \\ 
    & \Lr & \Z~\cite{Chernodub-17} & \Zt & \Zt \\
    & \Li & \Zt & $0$ & $2$\Z \\ \hline \hline
  \end{tabular}
\end{table}
%%%%%%%%%%

\clearpage
%%%%%%%%%%
\section{SIV.~Classification of previous works}

The previous works~\cite{Berry-04, Esaki-11, Zhen-15, Gonzalez-16, Leykam-17, Xu-17, Gonzalez-17, Chernodub-17, Zyuzin-18, Cerjan-18, Zhou-18-exp, Molina-18, Shen-18, Shen-Fu-18, Carlstrom-18, Moors-19, Okugawa-19, Budich-19, Yang-19, Zhou-19, Wang-19, Yoshida-19, Carlstrom-19, Zyuzin-19, Papaj-19, Kozii-17, Cerjan-18-exp, Luo-18, Lee-18-tidal, Chernodub-19, Bergholtz-19} on exceptional points and non-Hermitian topological semimetals are categorized into our classification (Table~\ref{tab: previous works}). For example, topologically stable exceptional points in two dimensions~\cite{Berry-04, Zhou-18-exp, Shen-18, Shen-Fu-18, Papaj-19, Kozii-17, Luo-18, Bergholtz-19} and exceptional rings in three dimensions~\cite{Berry-04, Xu-17, Gonzalez-16, Gonzalez-17, Zyuzin-18, Cerjan-18, Molina-18, Carlstrom-18, Yang-19, Carlstrom-19, Zyuzin-19, Cerjan-18-exp, Luo-18, Lee-18-tidal, Chernodub-19} correspond to the $\mathbb{Z}$ index for no symmetry, $p=2$, and point gap. Topologically stable exceptional rings in two-dimensional non-Hermitian systems with chiral symmetry~\cite{Esaki-11, Budich-19, Yoshida-19} ($\PT$ symmetry~\cite{Zhen-15, Okugawa-19, Zhou-19}) correspond to the $\mathbb{Z}$ ($\mathbb{Z}_{2}$) index for chiral ($\PT$) symmetry, $p=1$, and point gap. We note that an exceptional ring in two dimensions protected by pseudo-Hermiticity~\cite{Budich-19} (the Hermitian conjugate of $\mathcal{C}\mathcal{P}$ symmetry, or $\mathcal{P}\mathcal{T}_{-}$ symmetry~\cite{Okugawa-19}) is equivalent to that protected by chiral symmetry ($\mathcal{P}\mathcal{T}$ symmetry).

%\clearpage
%%%%%%%%%%
\section{SV.~Exceptional points and parity-time-symmetry breaking}

\begin{figure}[b]
\centering
\includegraphics[width=172mm]{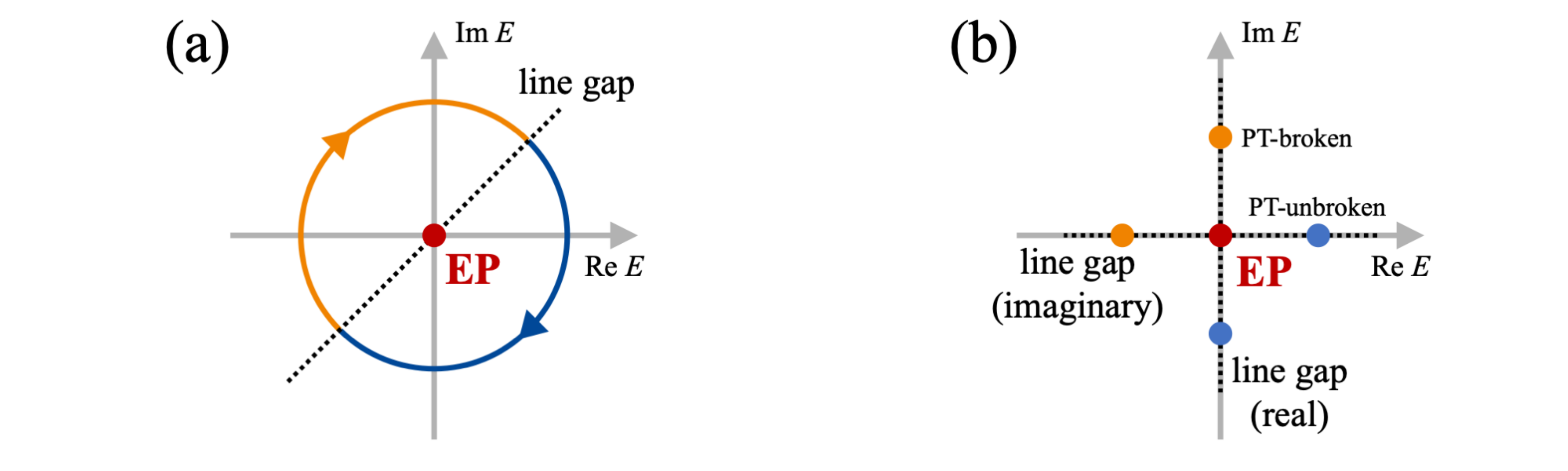} 
\caption{Exceptional points (EPs) and parity-time-symmetry breaking. In (a)~a two-dimensional system without symmetry and (b)~a one-dimensional system having $\PT$ symmetry with $(\PT) (\PT)^{*} = +1$, a point (line) gap is open (closed) around each EP (red point).}
	\label{fig: EP - PT}
\end{figure}

As discussed in the main text, a distinctive property of an exceptional point is swapping of eigenenergies and eigenstates upon its encircling [Fig.~\ref{fig: EP - PT}\,(a)]. Such an exceptional point indeed has the unique gapless structure with no Hermitian analogs around which a point gap is open but a line gap is closed [Fig.~\ref{fig: gapless}\,(d) in the main text]. Another typical exceptional points appear upon $\PT$-symmetry breaking~\cite{Heiss-12, Konotop-review, *Feng-review, *Christodoulides-review, *Alu-review, Bender-98, *Bender-02, *Bender-review} [see Eq.~(\ref{eq: PT symmetry}) in the main text for the definition of $\PT$ symmetry]. $\mathcal{PT}$-symmetric systems have entirely real spectra when $\mathcal{PT}$ symmetry is unbroken, whereas some eigenenergies form complex-conjugate pairs when it is spontaneously broken; between the two phases an exceptional point appears. We here consider a $\PT$-transition point ${\bm k} = {\bm k}_{\rm EP}$ at which two complex bands coalesce with each other and denote their right and left eigenstates as $\ket{\psi_{1, 2} \left( {\bm k} \right)}$ and $\langle\!\langle \psi_{1, 2} \left( {\bm k} \right) |$, normalized by $\langle\!\langle \psi_{m} \left( {\bm k} \right) |\psi_{n} \left( {\bm k} \right) \rangle = \delta_{mn}$ for ${\bm k} \neq {\bm k}_{\rm EP}$. When $\mathcal{PT}$ symmetry is unbroken and the spectrum is real, we have~\cite{Bender-98, *Bender-02, *Bender-review}
\begin{equation}
\mathcal{PT} \ket{\psi_{1} \left( {\bm k} \right)}^{*} = \ket{\psi_{1} \left( {\bm k} \right)},\quad
\mathcal{PT} \ket{\psi_{2} \left( {\bm k} \right)}^{*} = \ket{\psi_{2} \left( {\bm k} \right)}
\end{equation} 
with a proper choice of the gauges. On the other hand, when $\mathcal{PT}$ symmetry is spontaneously broken and the two eigenstates form a complex-conjugate pair, we have~\cite{Bender-98, *Bender-02, *Bender-review}
\begin{equation}
\mathcal{PT} \ket{\psi_{1} \left( {\bm k} \right)}^{*} = \ket{\psi_{2} \left( {\bm k} \right)}.
\end{equation} 
Since both equalities are compatible at the $\PT$-transition point ${\bm k} = {\bm k}_{\rm EP}$, the two bands coalesce, i.e.,
\begin{equation}
\ket{\psi_{1} \left( {\bm k}_{\rm EP} \right)} = \ket{\psi_{2} \left( {\bm k}_{\rm EP} \right)},
\end{equation} 
and thus form an exceptional point.

Remarkably, this $\mathcal{PT}$-transition-induced exceptional point can also have the unique gapless structure. For simplicity, we consider a $\PT$-symmetric system with $(\PT) (\PT)^{*} = + 1$ in one dimension. A representative model is given as 
\begin{equation}
H \left( k \right) = k \sigma_{x} + \ii \gamma \sigma_{z},\quad\PT = \sigma_{x},
\end{equation} 
which has exceptional points $k_{\rm EP} = \pm \gamma$. In fact, a point gap is open except for $k = k_{\rm EP}$, whereas a line gap for the imaginary (real) part is closed in the $\PT$-unbroken (-broken) phase [Fig.~\ref{fig: EP - PT}\,(b)]. Due to the point gap, the $\mathbb{Z}_{2}$ topological charge $\nu \in \{ 0, 1 \}$ can be defined by~\cite{Gong-18, Okugawa-19} 
\begin{equation}
\left( -1 \right)^{\nu} := \mathrm{sgn}\,\mathrm{det} \left[ H \left( k \right) - E \left( k_{\rm EP} \right) \right].
\end{equation} 
The change in $\nu$ signals $\PT$-symmetry breaking that accompanies an exceptional point. We note that this expression is applicable even if the degeneracy is not defective. Another expression for this $\mathbb{Z}_{2}$ topological charge is given by
\begin{equation}
\left( -1 \right)^{\nu} := \det w \left( k \right),
\end{equation}
where $w$ is a real matrix that satisfies $w_{mn} \left( k \right) := \langle\!\langle \psi_{m} \left( k \right) |\,\PT\,| \psi_{n} \left( k \right) \rangle^{*}$ and $w^{2} = ww^{*} = 1$. Since this charge takes $0$ ($1$) in the $\PT$-unbroken (broken) phase, its change signals $\PT$-symmetry breaking that accompanies an exceptional point.

%%%%%%%%%%
\section{SVI.~Recipe for exceptional points}

Based on our classification, we can systematically obtain non-Hermitian gapless systems having exceptional points:

\begin{enumerate}
\item In the classification table, choose a symmetry class that has nontrivial topology for a line gap in $p$ codimension.
\item Consider a 
$p$-dimensional Hermitian (or anti-Hermitian) gapless Dirac model $H \left( \bm{k} \right) = \sum_{i=1}^{p} v_{i} k_{i} \gamma_{i}$ that respects symmetry of the class and in which the matrix size $n$ of $\gamma_i$'s is minimal. The Dirac point at ${\bm k}=0$ is topologically stable due to the nontrivial topology for a line gap in $p$ codimension. Here note that topological charges for a line gap reduce to those of Hermitian systems. Since the energy spectrum is entirely real (or pure imaginary) and a line gap is closed at this Dirac point, it is located at zero energy (i.e., a reference point for a point gap in the complex-energy plane). Moreover, it is maximally degenerate ($n$-fold degenerate).

\item Add a non-Hermitian perturbation that preserves the symmetry and is independent of ${\bm k}$. The non-Hermiticity makes the spectrum complex-valued, and the Dirac point spreads in both momentum space and complex-energy plane. Because of the nontrivial topology for a line gap, there remains a gapless region with respect to a line gap, i.e., a finite segment of a reference line for the line gap. Then, if this gapless segment contains the reference point for a point gap, an exceptional point can appear at this reference point. Indeed, the line gap is closed but the point gap is open around the reference point, which is a characteristic of exceptional points. 

\item The obtained exceptional point has $p$ or less codimension, 
and its topological stability is ensured by the nontrivial topology for a point gap in the corresponding codimension. Reading off the possible stable structures (point, line, surface, and so on) and the topological invariants for a point gap from the classification table, we can confirm that it is indeed an exceptional point.

\item Very frequently, the symmetry-preserving non-Hermitian perturbation can be chosen so that the reference point keeps the $n$-fold degeneracy. In this case, we can easily prove that this reference point is defective (nondiagonalizable). In fact, since the perturbation is ${\bm k}$-independent, the perturbed energy spectrum passes the reference point at ${\bm k}={\bm k }_{\rm EP}\neq 0$. Then because $H \left( {\bm k}_{\rm EP} \right)$ is not proportional to the identity matrix, the $n$-fold degeneracy of $H \left({\bm k}_{\rm EP} \right)$ implies that $H \left({\bm k}_{\rm EP} \right)$ has a nontrivial Jordan form, i.e., $H \left( {\bm k}_{\rm EP} \right)$ is defective.

\end{enumerate}

A Weyl exceptional ring~\cite{Berry-04, Xu-17} and a topologically stable exceptional point in two dimensions~\cite{Zhou-18-exp, Shen-18} are examples that can be constructed with this recipe. In the absence of symmetry (i.e., class A), there appear a line-gapless point that is characterized by a $\mathbb{Z}$ topological charge (Chern number) in three dimensions ($p=3$), as well as a point-gapless point that is characterized by a $\mathbb{Z}$ topological charge (winding number) in two dimensions ($p=2$). Based on this fact, we begin with a Hermitian Weyl semimetal
\begin{equation}
H \left( {\bm k} \right) = k_{x} \sigma_{x} + k_{y} \sigma_{y} + k_{z} \sigma_{z},
\end{equation}
which has a robust degenerate point at ${\bm k} = 0$. We next add a non-Hermitian perturbation to it:
\begin{equation}
H \left( {\bm k} \right) = k_{x} \sigma_{x} + k_{y} \sigma_{y} + \left( k_{z} +\ii \gamma \right) \sigma_{z},\quad
E(\bm{k})=\sqrt{k_x^2+k_y^2+(k_z+\ii\gamma)^2}.\label{eq: nonh-Weyl}
\end{equation}
The Weyl point morphs into a ring of exceptional points at $k_{x}^{2} + k_{y}^{2} = \gamma^{2}$, $k_{z} = 0$~\cite{Berry-04, Xu-17}. After taking $k_{y}=0$ and regarding this non-Hermitian three-dimensional model as a two-dimensional one, we have
\begin{equation}
H \left( {\bm k} \right) = k_{x} \sigma_{x} +  \left( k_{z} +\ii \gamma \right) \sigma_{z},
\end{equation}
where a pair of exceptional points appears at $\left( k_{x}, k_{z} \right) = \left( \pm \gamma, 0 \right)$ in two dimensions~\cite{Zhou-18-exp, Shen-18}.

Another example is a chiral-symmetry-protected exceptional point in three dimensions discussed in the main text. In the presence of chiral symmetry (i.e., class AIII), there appear a line-gapless point that is characterized by a $\mathbb{Z}$ topological charge (winding number) in four dimensions ($p=4$), as well as a point-gapless point that is characterized by a $\mathbb{Z}$ topological charge (Chern number) in three dimensions ($p=3$). Based on this fact, we begin with a Hermitian gapless Dirac model with chiral symmetry in four dimensions:
\begin{equation}
H \left( {\bm k} \right) = k_{x} \sigma_{x} \tau_{x} + k_{y} \sigma_{x} \tau_{y} + k_{z} \sigma_{x} \tau_{z} + k_{w} \sigma_{y},
\end{equation}
which respects chiral symmetry with $\Gamma = \sigma_{z}$ [i.e., $\sigma_{z} H^{\dag} \left( {\bm k} \right) \sigma_{z}^{-1} = - H \left( {\bm k} \right)$]. Due to chiral symmetry, this model possesses a robust gapless point at ${\bm k} = 0$, which is topologically protected by the three-dimensional winding number $W=1$ defined as~\cite{Schnyder-08, *Ryu-10, Qi-10, Kawakami-18}
\begin{equation}
W := \int_{S^3} \frac{d^3k}{48\pi^2} \sum_{\alpha,\beta,\gamma=1}^4 \epsilon^{\alpha\beta\gamma}\,\mathrm{tr} \left[ \Gamma \left( H^{-1} \frac{\partial H}{\partial k_{\alpha}} \right) \left( H^{-1} \frac{\partial H}{\partial k_{\beta}} \right) \left( H^{-1} \frac{\partial H}{\partial k_{\gamma}} \right) \right],
\end{equation}
where $S^{3}$ is a hypersphere surrounding the gapless point ${\bm k} = 0$ and $\epsilon^{\alpha \beta \gamma}$ is the Levi-Civita symbol. Moreover, this gapless point is four-fold degenerate. Now we add a non-Hermitian perturbation to it:
\begin{equation}
H \left( {\bm k} \right) = k_{x} \sigma_{x} \tau_{x} + k_{y} \sigma_{x} \tau_{y} + k_{z} \sigma_{x} \tau_{z} + k_{w} \sigma_{y} + \ii \gamma \sigma_{z} \tau_{x},
\end{equation}
Noting $\{ \sigma_{x} \tau_{x}, \sigma_{z} \tau_{x} \} = \{ \sigma_{y}, \sigma_{z} \tau_{x} \} = 0$ and $[ \sigma_{x} \tau_{y}, \sigma_{z} \tau_{x} ] = [ \sigma_{x} \tau_{z}, \sigma_{z} \tau_{x} ] = 0$, we have
\begin{equation}
H^{2} \left( {\bm k} \right) = {\bm k}^{2} - \gamma^{2} + 2\ii \gamma \left( k_{y} \sigma_{x} \tau_{y} + k_{z} \sigma_{x} \tau_{z} \right),
\end{equation}
leading to the complex spectrum
\begin{equation}
E \left( {\bm k} \right) = \pm \sqrt{ {\bm k}^{2} - \gamma^{2} \pm 2\ii \gamma \sqrt{k_{y}^{2} + k_{z}^{2}}}\,.
\end{equation}
The original four-fold degenerate point thus morphs into an exceptional ring at $k_{x}^{2} + k_{w}^{2} = \gamma^{2}$, $k_{y} = k_{z} = 0$. After taking $k_{w}=0$ and regarding this non-Hermitian four-dimensional model as a three-dimensional one, we have
\begin{equation}
H \left( {\bm k} \right) = k_{x} \sigma_{x} \tau_{x} + k_{y} \sigma_{x} \tau_{y} + k_{z} \sigma_{x} \tau_{z} + \ii \gamma \sigma_{z} \tau_{x},
\end{equation}
where a pair of exceptional points appears at $\left( k_{x}, k_{y}, k_{z} \right) = \left( \pm \gamma, 0, 0 \right)$ in three dimensions. In fact, the Hamiltonian is defective (nondiagonalizable) at these points:
\begin{equation}
H \left( \pm \gamma, 0, 0 \right) = \gamma \left( \pm \sigma_{x} + \ii \sigma_{z} \right) \tau_{x},
\end{equation}
whose Jordan matrix $J$ is given as
\begin{equation}
J = \left( \begin{array}{@{\,}cccc@{\,}} 
	0 & 1 & 0 & 0 \\
	0 & 0 & 0 & 0 \\
	0 & 0 & 0 & 1 \\
	0 & 0 & 0 & 0 \\
	\end{array} \right).
\end{equation}

%%%%%%%%%%
\section{SVII.~Topological charge of a chiral-symmetry-protected gapless point}

In the presence of chiral symmetry defined by Eq.~(\ref{eq: chiral symmetry}) in the main text, i.e., 
\begin{equation}
\Gamma H^{\dag} \left( {\bm k} \right) \Gamma^{-1} = - H \left( {\bm k} \right),\quad
\Gamma^{2} = 1
\end{equation}
with a unitary matrix $\Gamma$, the topological charge of a non-Hermitian gapless point with odd $p \geq 1$ is given as the Chern number of the Hermitian matrix $\ii H \left( {\bm k} \right) \Gamma$ that is defined on a surface $S^{p-1}$ enclosing the gapless point~\cite{KSUS-18, Yoshida-19}. To see this fact, let us consider the following Hermitian matrix:
\begin{equation}
{\sf H} \left( {\bm k} \right) := 
\left( \begin{array}{@{\,}cc@{\,}} 
	0 & H \left( {\bm k} \right) \\
	H^{\dag} \left( {\bm k} \right) & 0 \\ 
	\end{array} \right),\quad
{\sf H}^{\dag} \left( {\bm k} \right) = {\sf H} \left( {\bm k} \right).
\end{equation}
Here the presence of a point gap for $H \left( {\bm k} \right)$ is equivalent to the presence of an energy gap for ${\sf H} \left( {\bm k} \right)$~\cite{Gong-18, KSUS-18}. Due to Eq.~(\ref{eq: chiral symmetry}) in the main text, ${\sf H} \left( {\bm k} \right)$ also respects chiral symmetry:
\begin{equation}
{\sf \Gamma}\,{\sf H} \left( {\bm k} \right) {\sf \Gamma}^{-1} 
= - {\sf H} \left( {\bm k} \right),\quad
{\sf \Gamma} := \left( \begin{array}{@{\,}cc@{\,}} 
	0 & \Gamma \\
	\Gamma & 0 \\ 
	\end{array} \right).
\end{equation}
Moreover, ${\sf H} \left( {\bm k} \right)$ respects intrinsic chiral symmetry by construction:
\begin{equation}
{\sf \Sigma}\,{\sf H} \left( {\bm k} \right) {\sf \Sigma}^{-1} 
= - {\sf H} \left( {\bm k} \right),\quad
{\sf \Sigma} := \left( \begin{array}{@{\,}cc@{\,}} 
	1 & 0 \\
	0 & -1 \\ 
	\end{array} \right).
\end{equation}
From the two chiral symmetries, we have the following commutation relation
\begin{equation}
\left[ {\sf H} \left( {\bm k} \right),\,\ii {\sf \Gamma \Sigma} \right] = 0.
\end{equation}
As a result, two Hermitian matrices ${\sf H} \left( {\bm k} \right)$ and $\ii {\sf \Gamma \Sigma}$ can be simultaneously diagonalized with a unitary matrix ${\sf U}$ as
\begin{equation}
{\sf U}^{\dag} {\sf H} \left( {\bm k} \right) {\sf U} = \left( \begin{array}{@{\,}cc@{\,}} 
	\ii H \left( {\bm k} \right) \Gamma & 0 \\
	0 & -\ii H \left( {\bm k} \right) \Gamma \\ 
	\end{array} \right),\quad
{\sf U}^{\dag} \left( \ii {\sf \Gamma \Sigma} \right) {\sf U} = \left( \begin{array}{@{\,}cc@{\,}} 
	1 & 0 \\
	0 & -1 \\ 
	\end{array} \right);\quad
{\sf U} := \frac{1}{\sqrt{2}} \left( \begin{array}{@{\,}cc@{\,}} 
	1 & -\ii \\
	\ii \Gamma & -\Gamma \\ 
	\end{array} \right).
\end{equation}
Therefore, the topology of non-Hermitian $H \left( {\bm k} \right)$ is captured by Hermitian $\ii H \left( {\bm k} \right) \Gamma$ with no symmetry, and the Chern number can be defined for $\ii H \left( {\bm k} \right) \Gamma$ on the surface $S^{p-1}$ with odd $p \geq 1$.

For the non-Hermitian model in Eq.~(\ref{eq: CS-p3}) in the main text, we have
\begin{eqnarray}
\ii H \left( {\bm k} \right) \Gamma
&=& k_{x} \sigma_{y} \tau_{x} + k_{y} \sigma_{y} \tau_{y} + k_{z} \sigma_{y} \tau_{z} - \gamma \tau_{x} \nonumber \\
&=& U \left( \begin{array}{@{\,}cc@{\,}} 
	\left( k_{x} -\gamma \right) \tau_{x} + k_{y} \tau_{y} + k_{z} \tau_{z} & 0 \\
	0 & - \left( k_{x} + \gamma \right) \tau_{x} - k_{y} \tau_{y} - k_{z} \tau_{z} \\ 
	\end{array} \right) U^{\dag},
\end{eqnarray}
where a $SU \left( 2 \right)$ rotation $U := e^{-\ii \left( \pi/4 \right) \sigma_{x}}$ changes $\sigma_{y}$ to $\sigma_{z}$ (i.e., $U\sigma_{y}U^{\dag} = \sigma_{z}$). Since this expression is a direct sum of the two Weyl semimetals whose gapless points are located at ${\bm k} = {\bm k}_{\rm EP} = \left( \pm \gamma, 0, 0 \right)$, we have the Chern number $\pm 1$ for each exceptional point ${\bm k}_{\rm EP}$.

%\bibliography{NH-EP-TSM}

\end{document}